\documentclass[twocolumn]{aastex63}
\usepackage{comment}
\usepackage{amsmath}
\usepackage{rotating}
\usepackage{layouts}
\usepackage{tasks}
\usepackage{enumitem}
\usepackage{placeins}
\usepackage{xcolor}
\usepackage{soul}

\shorttitle{C/O Meta Analysis}
\shortauthors{Hoch et al.}

\begin{document}

\title{Assessing the C/O Ratio Formation Diagnostic: A Potential Trend with Companion Mass}

\author[0000-0002-9803-8255]{Kielan K. W. Hoch}
\affiliation{Center for Astrophysics and Space Sciences,  University of California, San Diego, La Jolla, CA 92093, USA}
\affiliation{Space Telescope Science Institute, 3700 San Martin Dr, Baltimore, MD 21218, USA}

\author[0000-0002-9936-6285]{Quinn M. Konopacky}
\affiliation{Center for Astrophysics and Space Sciences,  University of California, San Diego, La Jolla, CA 92093, USA}

\author[0000-0002-9807-5435]{Christopher A. Theissen}
\altaffiliation{NASA Sagan Fellow}
\affiliation{Center for Astrophysics and Space Sciences, University of California, San Diego, La Jolla, CA 92093, USA}

\author[0000-0003-2233-4821]{Jean-Baptiste Ruffio}
\affiliation{Center for Astrophysics and Space Sciences,  University of California, San Diego, La Jolla, CA 92093, USA}

\author[0000-0002-7129-3002]{Travis S. Barman}
\affiliation{Lunar and Planetary Laboratory, University of Arizona, Tucson, AZ 85721, USA}

\author[0000-0003-4203-9715]{Emily L. Rickman}
\affiliation{European Space Agency (ESA), ESA Office, Space Telescope Science Institute, Baltimore, MD 21218, USA}


\author[0000-0002-3191-8151]{Marshall D. Perrin}
\affiliation{Space Telescope Science Institute, 3700 San Martin Dr, Baltimore, MD 21218, USA}

\author[0000-0003-1212-7538]{Bruce Macintosh}
\affiliation{Kavli Institute for Particle Astrophysics and Cosmology, Stanford University, Stanford, CA 94305, USA}

\author[0000-0002-4164-4182]{Christian Marois}
\affiliation{NRC Herzberg Astronomy and Astrophysics, 5071 West Saanich Rd, Victoria, BC V9E 2E7, Canada}

\correspondingauthor{Kielan K. W. Hoch}
\email{khoch@stsci.edu}


\begin{abstract}
    The carbon-to-oxygen (C/O) ratio in an exoplanet atmosphere has been suggested as a potential diagnostic of planet formation. Now that a number of exoplanets have measured C/O ratios, it is possible to examine this diagnostic at a population level. Here, we present an analysis of currently measured C/O ratios of directly imaged and transit/eclipse planets. First, we derive atmospheric parameters for the substellar companion HD 284149 AB b using data taken with the OSIRIS integral field spectrograph at the W.M. Keck Observatory and report two non-detections from our ongoing imaging spectroscopy survey with Keck/OSIRIS. We find an effective temperature of $T_\mathrm{eff} = 2502$ K, with a range of 2291--2624 K, $\log g=4.52$, with a range of 4.38--4.91, and [M/H] = 0.37, with a range of 0.10--0.55. We derive a C/O of 0.59$^{+0.15}_{-0.30}$ for HD 284149 AB b. We add this measurement to the list of C/O ratios for directly imaged planets and compare them with those from a sample of transit/eclipse planets. We also derive the first dynamical mass estimate for HD 284149 AB b, finding a mass of $\sim$28 $M_\mathrm{Jup}$. There is a trend in C/O ratio with companion mass ($M_{\mathrm{Jup}}$), with a break seen around 4\,$M_{\mathrm{Jup}}$. We run a Kolmogorov-Smirnov and an Anderson-Darling test on planets above and below this mass boundary, and find that they are two distinct populations. This could be additional evidence of two distinct populations possibly having two different formation pathways, with companion mass as an indicator of most likely formation scenario.
\end{abstract}
\section{Introduction}


Over 5,000 exoplanets are now known, and only about 80 of these objects have atmospheric data from the Hubble Space Telescope (HST), the Spitzer Space Telescope, and/or various ground-based observatories. There are a few main methods to obtain atmospheric information on these companions, such as direct imaging and the transit or eclipse techniques \citep{sing2016,tsiaras2018,pinhas2019,mansfield2021}. Direct imaging surveys have revealed a unique population of massive, widely separated exoplanets \citep{rameau2013,biller2013,galicher2016,nielsen2019,vigan2021}. Revealing these gas giant companions has inspired a number of novel and complex theories to explain their large sizes and separations, such as core/pebble accretion \citep{johansen2017}, dynamical scattering \citep{rasio1996,weidenschilling1996,chatterjee2008}, disk instability \citep{kuiper1951,cameron1978}, and cloud fragmentation \citep{toomre1964}. These objects still remain a mystery for many of in situ formation mechanisms. 

Most transiting or eclipsing gas giant planets with spectral information are ``Hot Jupiters" with temperatures around 1000--2500 K.  They orbit close to their host stars ($<$1 au), and can have a variety of masses spanning from a percent of a Jupiter mass to 30 Jupiter masses. They are thought to form through a three step process starting from core accretion, then runaway gas accretion to form the atmosphere, and finally inward migration to replicate what we see today \citep{mizuno1980,bodenheimer1986,ikoma2000}. The core accretion step is thought to occur in beyond the ice line of a protoplanetary disk where an abundance of solid material leads to rapid growth of a planetary core before gas dispersal in the disk. With this scenario in mind, the composition of these exoplanets should be sub-stellar in carbon and oxygen because they would be sequestered in the solid core. However, measured bulk metallicities of Jupiter, Saturn, and some exoplanets are shown to be super-stellar, which complicates tracing the formation scenario from composition of the atmospheres \citep{li2020,wong2004,visscher2005,madhusudhan2011}.

Relative abundance measurements of molecules were first done on transiting Hot Jupiters to characterize their atmospheric temperature and pressure structures \citep{fortney2010}. In particular, carbon-to-oxygen ratios (C/O) were measured and they differed from their host star significantly, furthering the need for a possible explanation \cite[e.g. WASP-12b, C/O $\approx 1$;][]{madhu2011}. \cite{oberg2011} engineered a framework to use the C/O ratio to trace how and where the object formed in the protoplanetary disk. In summary, a stellar C/O ratio is expected for companions forming through gravitational instabilities in the disk, where all material is mixed, and for planets forming interior to both the water snow-line and the carbon-grain evaporation line. Superstellar C/O ratios could indicate atmosphere formation from mainly gas accretion outside of the water snowline, possibly via core accretion. Other studies also connected the C/O ratio to formation mechanisms and evolution in the protoplanetary disk \citep{madhu2012,alidib2014,madhu2014c,thiabaud2015}. Many C/O ratios were measured for transiting planets along with studies to explain some of the higher ratios \cite[i.e.,][]{madhu2011b,madhu2012,madhu2012b,moses2013,teske2013,teske2013b,line2014}. Directly imaged companions also started to have their C/O ratios measured, with the hope that these youthful systems would not have undergone significant chemical evolution \citep{konopacky2013,wilcomb2020,molliere2020,petrus2021,zhang2021,ruffio2021,hoch2022,palma-bifani2022}. The C/O ratio measurement may be able to trace the formation location of these objects and how they collected gas and solids in their evolving protoplanetary disks \citep{oberg2011,madhu2017,mordasini2016,booth2017,cridland2019,eistrup2018}. More recently, \cite{molliere2022} conducted a study using state-of-the-art formation models, including the chemical evolution of the protoplanetary disk, showing that chemical abundance measurements are crucial for informing new formation models. 


Spectral characterization of exoplanets and substellar companions can shed light on how they may have formed. The current directly imaged companions are young ($<$ 200 Myr), and thus their atmospheres have not undergone extensive post-formation chemical evolution and could point towards how and where they formed in the protoplanetary disk. In contrast, the transiting Hot Jupiters are typically older and have gone through significant chemical evolution. Still, atmospheric composition measurements may provide hints about their formation. 

Here, we present a statistical analysis of the C/O ratio formation tracer using compiled values from directly imaged companions and a sample of transit/eclipse companions. In Section \ref{sec:c/o_ratio}, we report the C/O ratios used for this work and how they were measured. In Section\ref{sec:data3}, we report our observations of HD 284149 AB b, data reduction methods, atmospheric modeling, C/O ratio measurement, and non-detections of two targets with the OSIRIS instrument on the W.M. Keck Telescope. In section \ref{sec:dyn_mass} we measure dynamical masses for HD 284149 AB b and HD 284149 B. In Section \ref{sec:c-o_analysis} we compare C/O ratios of 25 transiting and eclipsing planets from \cite{changeat2022} to C/O ratios of the directly imaged planets. We find a trend between C/O ratio and companion mass and compute two metrics to determine if the planets originate from the same underlying population. In Section \ref{sec:discussion3} we discuss the implications of our results and in Section \ref{sec:conclusions} we discuss our conclusions and future work. 


\section{C/O Ratio Measurements from the Literature}
\label{sec:c/o_ratio}
\subsection{Direct Imaging Spectroscopy}
Direct imaging spectroscopy has allowed for the C/O ratio measurements of many imaged substellar companions.  Here we review the current suite of C/O ratios derived for directly imaged planets.  \cite{lee2013} ran a retrieval on low-resolution spectral data and photometric points on HR 8799 b from \cite{barman2011}, \cite{currie2011}, \cite{skemer2012}, and \cite{galicher2011} resulting in C/O ratios that were approximately unity across their various models. \cite{lavie2017} created an open source code HELIOS-RETRIEVAL and ran this code on the HR 8799 planets using data from \cite{zurlo2016}, \cite{ingraham2014}, \cite{barman2011}, and \cite{barman2015}. They estimated C/O values that ranged from superstellar for HR 8799 b to consistent with stellar for HR 8799 c and substellar for HR 8799 d and e. \cite{molliere2020} derived a C/O ratio of 0.60$^{+0.07}_{-0.08}$ for HR 8799 e through atmospheric retrievals using their code \textit{petitRADTRANS} \citep{molliere2019} with the added effect of multiple scattering to better treat cloudy objects. They ran retrievals on $K$-band GRAVITY data \citep{gravity2017} and archival SPHERE and GPI data. \cite{gravity2020} measured C/O ratios for $\beta$ Pic b using forward modeling with the newest ExoREM grid \cite{charnay2018} (0.43$\pm$0.05) and retrievals with \textit{petitRADTRANS} (0.43$^{+0.04}_{-0.03}$) on $K$-band GRAVITY spectra and GPI $Y$-$J$-$H$- band spectra. \cite{petrus2021} calculated an upper limit for the C/O ratio of HIP 65426 b to be $\le$0.55 using a Bayesian inference with BT-Settl with the ForMoSA code on SINFONI medium-resolution $K$-band data \citep{petrus2020}. A C/O ratio for TYC 8998-760-1 b (YSES-1 b) was derived to be 0.52$^{+0.04}_{-0.03}$ by \cite{zhang2021} using Bayesian retrieval analysis on SINFONI medium-resolution $K$-band data with \textit{petitRADTRANS}.  \cite{palma-bifani2022} forward modeled $J$-, $H$-, and $K$-band SINFONI data of AB Pic b using  ExoREM and BT-SETTL13 atmospheric models and the ForMoSA forward modeling code to obtain a C/O ratio of 0.58$\pm$0.08. \cite{brown2023} re-reduced SPHERE data on 51 Eri b and performed a retrieval using \textit{petitRADTRANS} on low-resolution Y-H spectrum and revised K1-K2 photometry to obtain a C/O ratio of 0.38$\pm$0.09. \cite{wang2023} performed an atmospheric retrieval on HR 8799 c photometric data along with low- and high-resolution spectroscopic data (R$\sim$20-35,000) and obtained a C/O ratio of 0.67$^{+0.12}_{-0.15}$. 

Our team has derived C/O ratios for six directly imaged companions using the OSIRIS IFU. C/O ratios for the three HR 8799 planets, HR 8799 b, c, and d were derived by \cite{ruffio2021} to be 0.578$^{+0.004}_{-0.005}$, 0.562$\pm0.004$, and 0.551$^{+0.005}_{-0.004}$ respectively quoting statistical uncertainties.  Notably, our values for the HR 8799 planets are consistent with those from other high-resolution spectra \citep{wang2023}, but not with all those derived from retrievals of low-resolution spectra. When using the values in \cite{ruffio2021} we adopt the C/O uncertainties calculated in previous measurements by \citet{konopacky2013} of 0.1 instead, because these measurements are dominated by model uncertainties that are not captured during the fitting process.  \cite{ruffio2021} used a custom \textit{PHOENIX} model grid and a forward modeling approach on OSIRIS IFU data to derive these ratios. \cite{wilcomb2020} derived the C/O ratio for $\kappa$ And b to be 0.704$^{+0.09}_{-0.24}$ following a forward modeling approach on the OSIRIS data using a custom \textit{PHOENIX} model grid. We measured VHS 1256 b's C/O ratio as 0.590$^{+0.28}_{-0.354}$ in \cite{hoch2022} following the same approach as \cite{wilcomb2020} using custom \textit{PHOENIX} models that included a cloud parameter to treat the thick clouds of an ``L/T" transition object. \cite{petrus2023} measured the upper limit of the C/O ratio for VHS 1256 b to be $>$0.63, which is encompassed in the error bars in \cite{hoch2022}.


\subsection{Transit/Eclipse Spectroscopy}
Transiting exoplanets can occasionally offer low- to high-resolution spectra that can, in turn, allow for the derivation of atmospheric properties such as the C/O ratio. Low-resolution data cannot reveal individual spectral lines, unlike moderate resolution data, so retrievals are often used to derive these properties. \cite{changeat2022} reanalyzed HST WFC3 G141 Grism data using their pipeline, \textit{Iraclis} \citep{tsiaras2016}, and Spitzer data of 25 Hot Jupiters.  The retrieval code used was \textit{Alfnoor} \citep{changeat2020}, and the equilibrium chemistry retrievals were conducted using the GGChem code \citep{woitke2018}. Furthermore, \cite{changeat2022} used the two retrieval methods on both HST and Spitzer data and on them individually to obtain C/O ratio measurements for all 25 exoplanets in their sample. Some high-resolution spectra of transiting planets have been obtained using instruments such as CFHT/SPIRou, Keck/KPIC, and GG/IGRINS to better constrain the C/O ratio and individual chemical abundances of Hot Jupiters \citep{boucher2023,finnerty2023,pelletier2021,line2021}. 

\subsection{System Parameter Compilation}
We chose to use C/O ratio values obtained using moderate- high-resolution data for the directly imaged companions.  The inconsistency of the moderate/high resolution C/O values with those from low resolution retrievals is something of a conundrum.  However, the ability to resolve spectral lines in higher resolution data is advantageous.  There are instances of retrievals returning C/O ratios near unity that are highly inconsistent with the moderate resolution spectra, which would show different spectral lines near C/O$\sim$1. Therefore, we choose to use results from data where spectral lines are resolved. This results in not all directly imaged planets with C/O estimates described in the previous section being included in our analysis below, but we believe offers a sample that is the most consistent.  For the most part only low resolution data is available for transiting planets, and so our goal for this population  is uniformity in approach and analysis.  We therefore adopt the C/O ratios from \cite{changeat2022}, which presents a uniform sample and analysis of transiting Hot Jupiters. 

We compiled various parameters of both the directly imaged and transiting systems, choosing the most recent reported values, such as projected separation (au), companion mass ($M_{\mathrm{Jup}}$), host star mass ($M_{\odot}$), and age (Myr) in addition to C/O ratios of the respective planets. For the directly imaged planets, we used dynamical masses when there were measurements available. For the transiting planets, we used values computed by \cite{changeat2022}. In particular, we chose \cite{changeat2022} C/O ratio values that were derived through equilibrium chemistry retrieval (eq) on the combination of Spitzer and HST data. If the object did not have a measurement from the retrieval on both data-sets, we chose the HST-data only measurement from the equilibrium retrieval. This provided us with two populations, directly imaged planets and transit/eclipse planets, with derived C/O ratios that we could conduct a population comparison.  Our compiled values are presented in Table \ref{tab:megatable}. We chose the values from \cite{changeat2022} because they were all analyzed in the same way, by the same code, rather than obtaining values from literature that may be plagued with different systematics.

\section{Additional Data from OSIRIS} \label{sec:data3}

As described in the previous section, a significant number of the sources with C/O ratios compiled here come from our previous work on OSIRIS \citep{barman2015,konopacky2013,ruffio2021,wilcomb2020,hoch2022}.  Here we report on one additional observation that we add to the full list of sources.  We also describe two non-detections in our OSIRIS survey.

\subsection{Target Information}
HD~284149~AB~b is a substellar companion orbiting at $\sim$3.6$\arcsec$ from the F8 star HD 284149 A and a stellar companion HD 284149 AB b discovered by \cite{bonavita2014,bonavita2017}. The HD~284149 system has been proposed to be a part of the Taurus-Ext association \citep{luhman2017,kraus2017,daemgen2015} with an age estimate by \cite{bonavita2014} of 25$^{+25}_{-10}$ Myr. \cite{bonavita2017} found an effective temperature of 2395 $\pm$ 113 K, a spectral type of M9 $\pm$ 1, and a mass of 26 $\pm$ 3~M$_{\mathrm{Jup}}$ for the substellar companion HD~284149~AB~b. We calculate dynamical masses for components of the HD~284149 system that provide updated model-independent mass values of $170^{+23}_{-16}~M_{\rm{Jup}}$ (or $0.16\pm0.02~M_{\odot}$) and $28.26^{+0.75}_{-1.0}~M_{\rm{Jup}}$ HD~284148~B and HD~284149~AB~b respectively as described in Section~\ref{sec:dyn_mass}.

\subsection{Data Reduction}
\begin{figure*}
\centering
  \includegraphics[width=3cm]{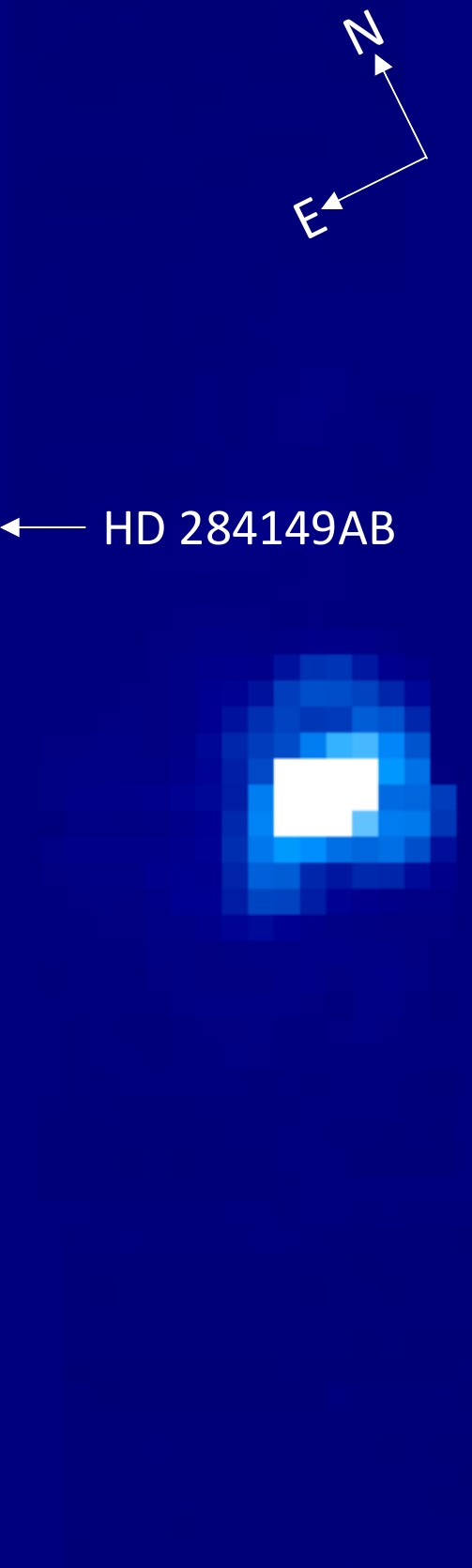}
\caption{An example data cube image, collapsed in wavelength via median, from our OSIRIS HD 284149 AB b dataset. The field of view is $0.32\arcsec$ x $1.28\arcsec$. The bright spot shows the companion clearly visible without any need for speckle removal.  }
\label{fig:osiris3}
\end{figure*}

HD 284149 AB b was observed on 2017 November 3 with the OSIRIS integral field spectrograph (IFS, \citealt{larkin2006}) on the W.M. Keck I telescope. We used the $K$ broadband mode (1.965--2.381 $\mu$m) with a spatial sampling of 20 milliarcseconds per lenslet. We integrated on the object for 50 minutes via 5 exposures of 10 minutes each, where we dithered up and down by a 2-3 pixels between exposures. Observations of a blank patch of sky and of our A0V telluric standard (HIP 16095) were obtained close in time with the object data. We also obtained dark frames with exposure times matching our dataset. The data were reduced using the OSIRIS data reduction pipeline \citep[DRP;][]{krabbe2004,lockhart2019}. Data cubes (x, y, $\lambda$) were generated using the OSIRIS DRP with rectification matrices provided by the observatory for the correct time frame of the observations. At the advice of the DRP working group, we did not use the Clean Cosmic Rays DRP module (T. Do, priv. comm) and we did not use scaled sky subtraction. An example data cube image frame is shown in Figure \ref{fig:osiris3}.

Once the one-dimensional spectra for the telluric sources were extracted, we used the DRP to remove hydrogen lines, divide by a blackbody spectrum of the appropriate temperature, and combine all standard star spectra. The individual telluric spectra were median combined and produced a telluric calibrator spectrum. The telluric correction for HD 284149 AB b was then completed by dividing the final telluric calibrator spectrum from all object frames.

Once the object data cubes are fully reduced, we identify the location of the target. When conducting high contrast observations with OSIRIS, the location of a companion can be challenging to find due to the brightness of the speckles from the host star. For HD 284149 AB b, the separation from the host star is wide enough and the companion is bright enough that the speckles from the host star do not impact the spectrum of the companion, and identification is straightforward (see Figure \ref{fig:osiris3}). 

\begin{figure*}
\centering
  \includegraphics[width=\textwidth]{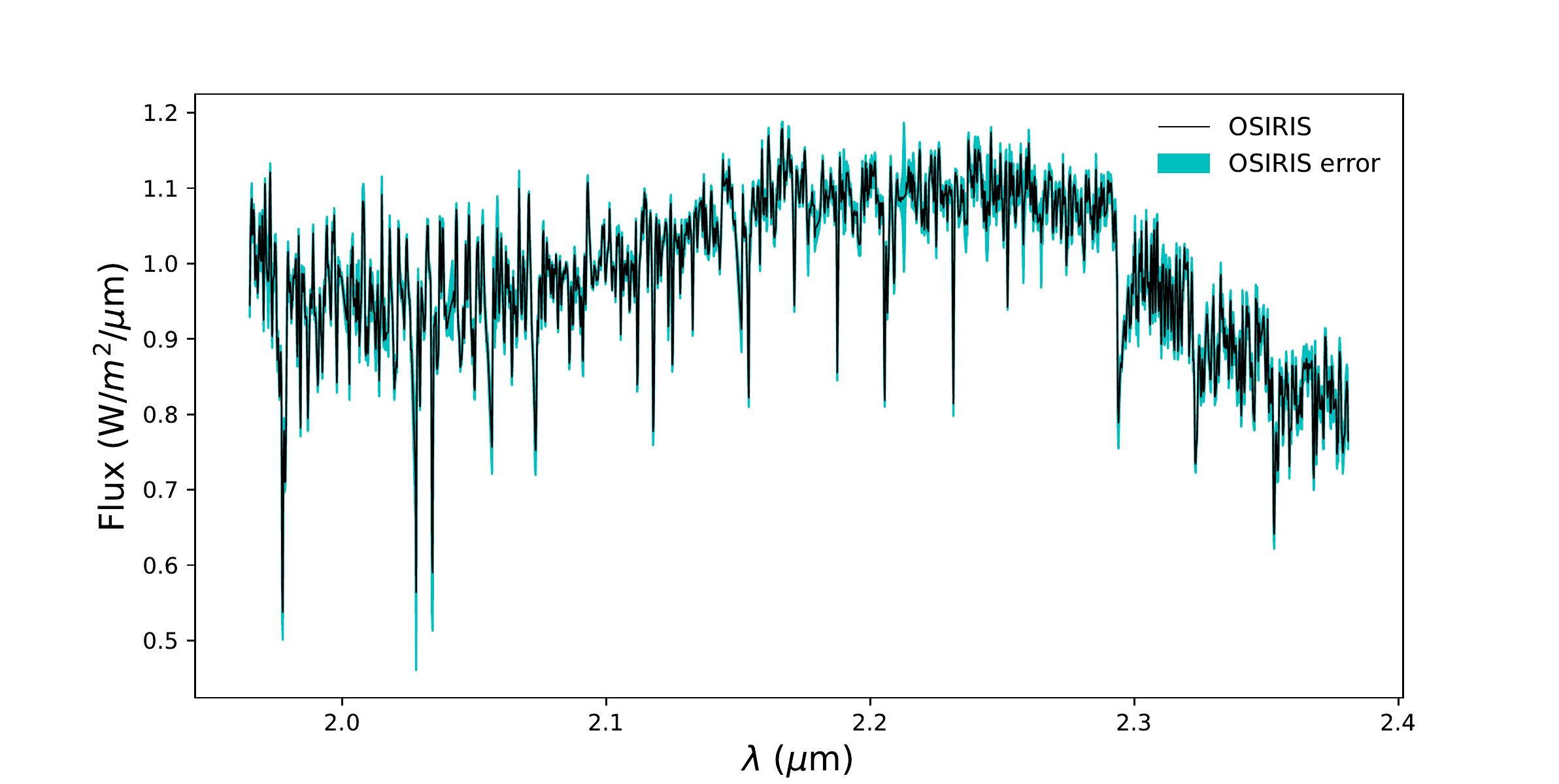}
\caption{Our fully reduced, combined, and flux calibrated, moderate-resolution OSIRIS $K$-band spectra of HD 284149 AB b. The errors are calculated from the RMS of the individual spectra at each wavelength. The error includes uncertainties in both the continuum and the lines. The total spectral uncertainties are represented as a shaded blue region. The CO bandhead is distinctly visible at 2.3 $\mu$m.}
\label{fig:hd284_flam}
\end{figure*}

We extract the object spectrum from the data cubes using a box of $3 \times 3$ spatial pixels (spaxels). Once we extracted the HD 284149 AB b spectra from each frame for all data cubes, we then normalize each individual spectrum to account for PSF and background fluctuations. Finally, we median-combine all 5 individual spectra, and do a barycentric correction on the individual spectra. To calibrate the flux of our spectra we calculated the flux at each wavelength such that, when integrated, the flux matches the most recent $K$-band apparent magnitude ($14.332 \pm 0.04$) from \cite{bonavita2014}.

Uncertainties were calculated by determining the RMS between the individual spectra at each wavelength. These uncertainties include contributions from statistical error in the flux of the planet and the molecular lines. The OH sky lines are subtracted extremely well and have a negligible contribution to the uncertainties. The final combined and flux calibrated spectrum is shown in Figure \ref{fig:hd284_flam}.

Narrow spectral features are more easily analyzed by removing the continuum because we can avoid low spatial frequency errors (from residual and/or faint speckles) that affect the continuum shape. We remove the continuum from our fully reduced and flux calibrated spectra using a similar continuum removal process we have used in the past for OSIRIS data (e.g., \citealt{barman2015,wilcomb2020,hoch2022}). To remove the continuum, we employ a high-pass filter with a kernel size of 200 spectral bins to each of the individual spectra. We then subtract the smoothed spectrum without spectral lines from the original spectra. Once all the individual spectra are flattened, we median combine them using the method for the continuum spectra and find the uncertainties by determining the RMS of the individual spectra at each wavelength.  

\subsection{Spectral Modeling} \label{sec:modeling3}
To determine the effective temperature ($T_\mathrm{eff}$), surface gravity ($\log g$), and metallicity ([M/H]) of HD 284149 AB b we used the \textit{PHOENIX} model-based G\"ottingen spectral library \citep{husser2013}. We chose to use the G\"ottingen spectral library because they are \textit{PHOENIX} models that cover the estimated temperature and surface gravity of HD 284149 AB b from previous studies, and therefore there was no need to create additional custom \textit{PHOENIX} models. 

Here, we use a forward-modeling process following \cite{blake2010}, \cite{burgasser2016}, \cite{hsu2021}, and \cite{theissen2022} to determine the best-fit model from the grid. The effective temperature ($T_\mathrm{eff}$), surface gravity ($\log g$), and metallicity ([M/H]) are inferred using a Markov Chain Monte Carlo (MCMC) method built on the \texttt{emcee} package that uses an implementation of the affine-invariant ensemble sampler \citep{goodman2010,foreman-mackey2013}. Our MCMC runs used 100 walkers, 500 steps, and a burn-in of 400 steps to ensure parameters were well mixed,  and the description of the MCMC calculations are described in \cite{wilcomb2020}.

\begin{figure*}
\centering
  \includegraphics[width=\textwidth]{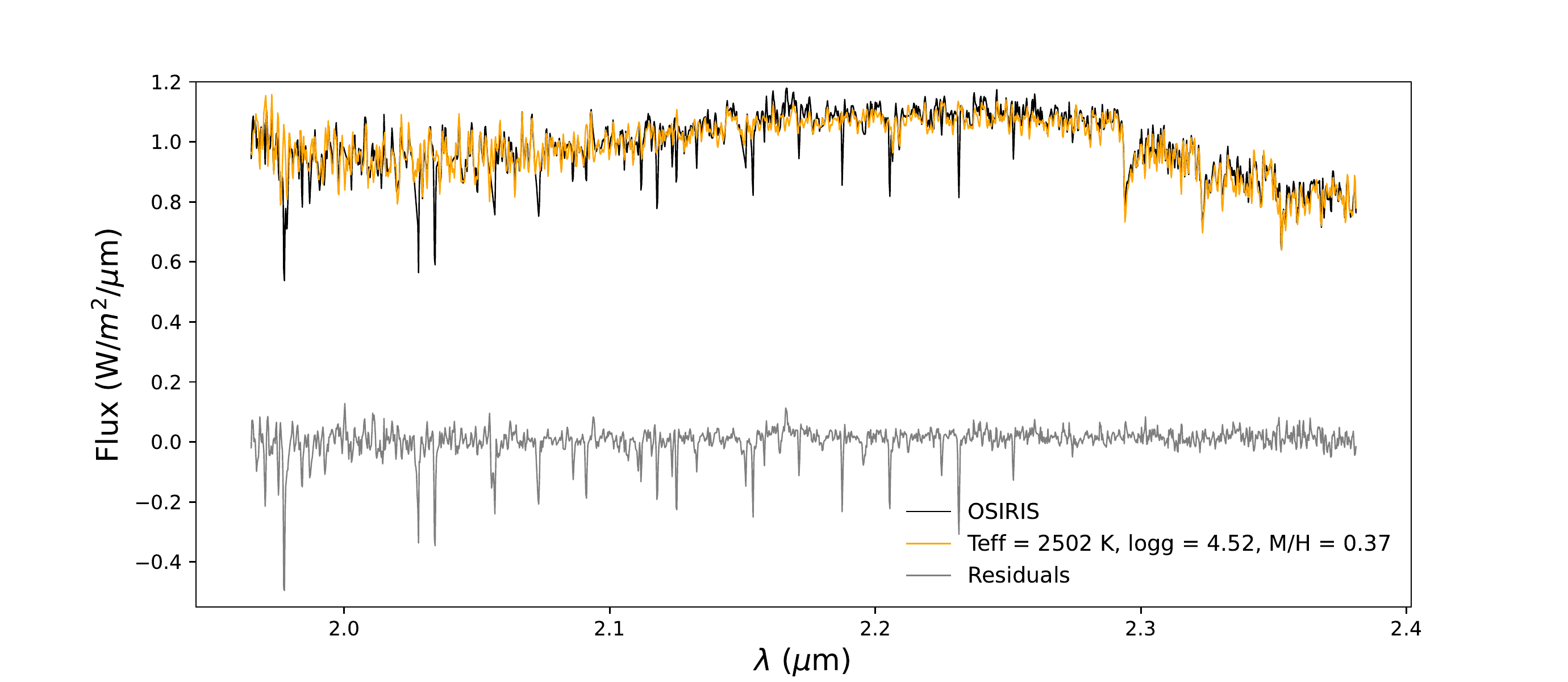}
\caption{Our fully reduced, combined, and flux calibrated moderate-resolution OSIRIS $K$-band spectra of HD 284149 AB b in black plotted against the best-fit model in orange. The residuals are plotted below in gray.}
\label{fig:hd284_cont_fit}
\end{figure*}

\begin{figure*}
\centering
  \includegraphics[width=\textwidth]{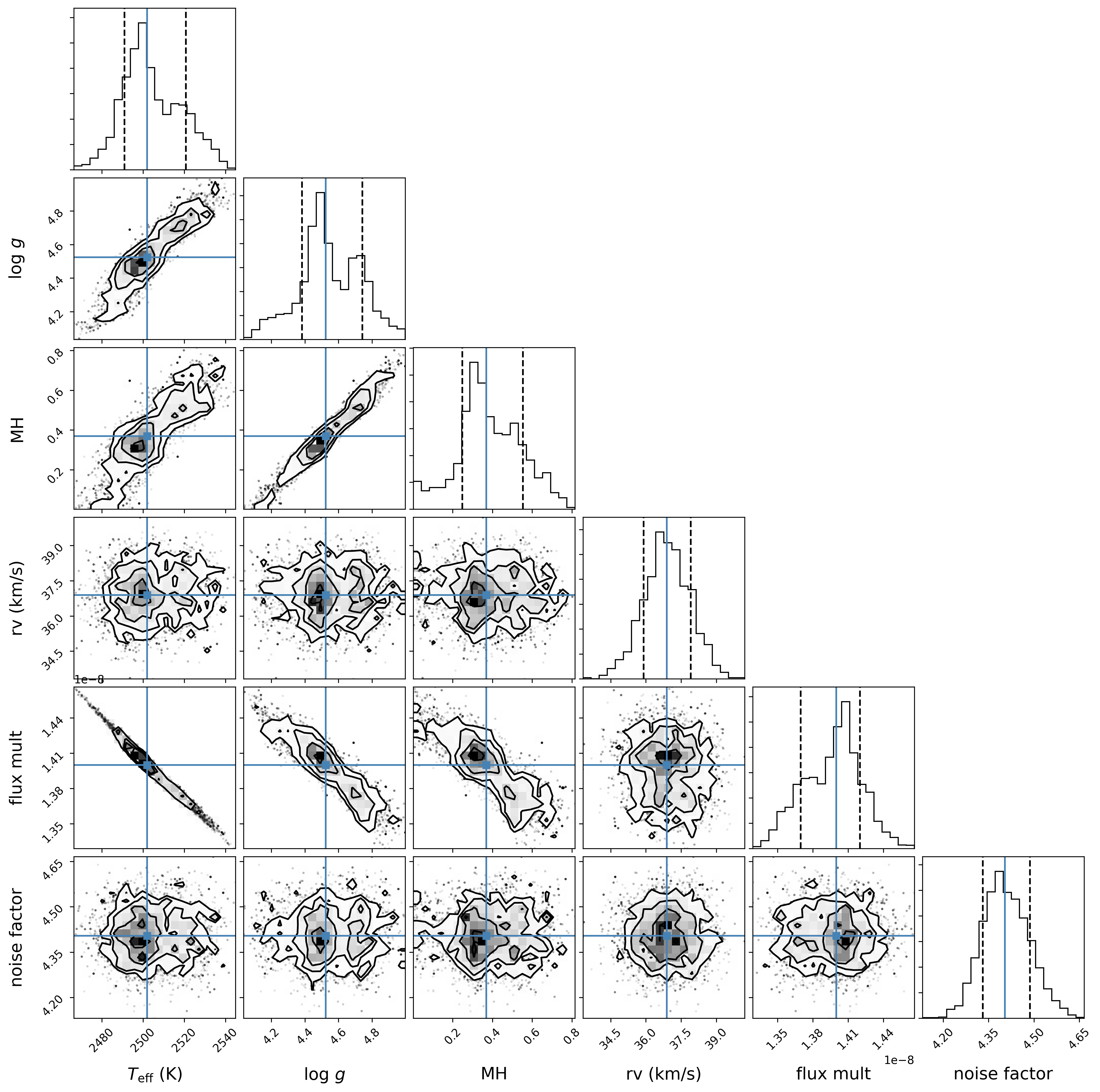}
\caption{Corner plot from our MCMC fits for our continuum OSIRIS $K$-band spectra. The diagonal shows the marginalized posteriors. The covariances between all the parameters are in the corresponding 2-d histograms. The blue lines represent the 50th percentile, and the dotted lines represent the 16 and 84 percentiles. The ``flux mult" corresponds to the dilution factor that scales the model by \((radius)^2 (distance)^{-2}\). }
\label{fig:hd284_cont_corner}
\end{figure*}

The results show that HD 284149 AB b has a super-solar metallicity.  \cite{dorazi2011} derived that the Taurus-Auriga association has a mean solar metallicity ([Fe/H]=-0.01$\pm$0.05). If this value is adopted for the Taurus-Ext association, the substellar companion is enhanced comparatively. The best fit parameters from our continuum $K$-band OSIRIS data are $T_\mathrm{eff}$ = 2502$^{+19}_{-11}$ K, $\log g$ = 4.52$^{+0.22}_{-0.14}$, and [M/H] = 0.37$^{+0.18}_{-0.12}$. The temperature results are in good agreement with the temperature from \cite{bonavita2017}, which was 2395 $\pm$ 113 K. Figures \ref{fig:hd284_cont_fit}--\ref{fig:hd284_cont_corner} show the best-fit model for the continuum data and the corresponding corner plot from our MCMC analysis.   We note that there are several lines that appear to be stronger in the data than in the model in Figure \ref{fig:hd284_cont_fit}.  We have verified that these are not residual telluric contamination, but instead real lines, likely from scandium, yttrium, titanium, and iron.  We note that stronger lines in the data than the $PHOENIX$ models is a known phenomenon for scandium, yttrium, and titanium in high-metallicity, low temperature objects.  It is due to the stronger impact of non-LTE effects on these lines due to hyperfine splitting (e.g., \citealt{Bergemann2012,Thorsbro2018}).  Since our focus in this work is carbon and oxygen, and the mismatch of these lines will not change the derived effective temperature or log(g), which are strongly dependent on the continuum shape and CO lines, we do not attempt further modeling of these lines.  Given the good signal-to-noise ratio of this spectrum, potential future work could involve detailed modeling of these lines to obtain precise abundances for these elements.

\begin{figure*}
\centering
  \includegraphics[width=\textwidth]{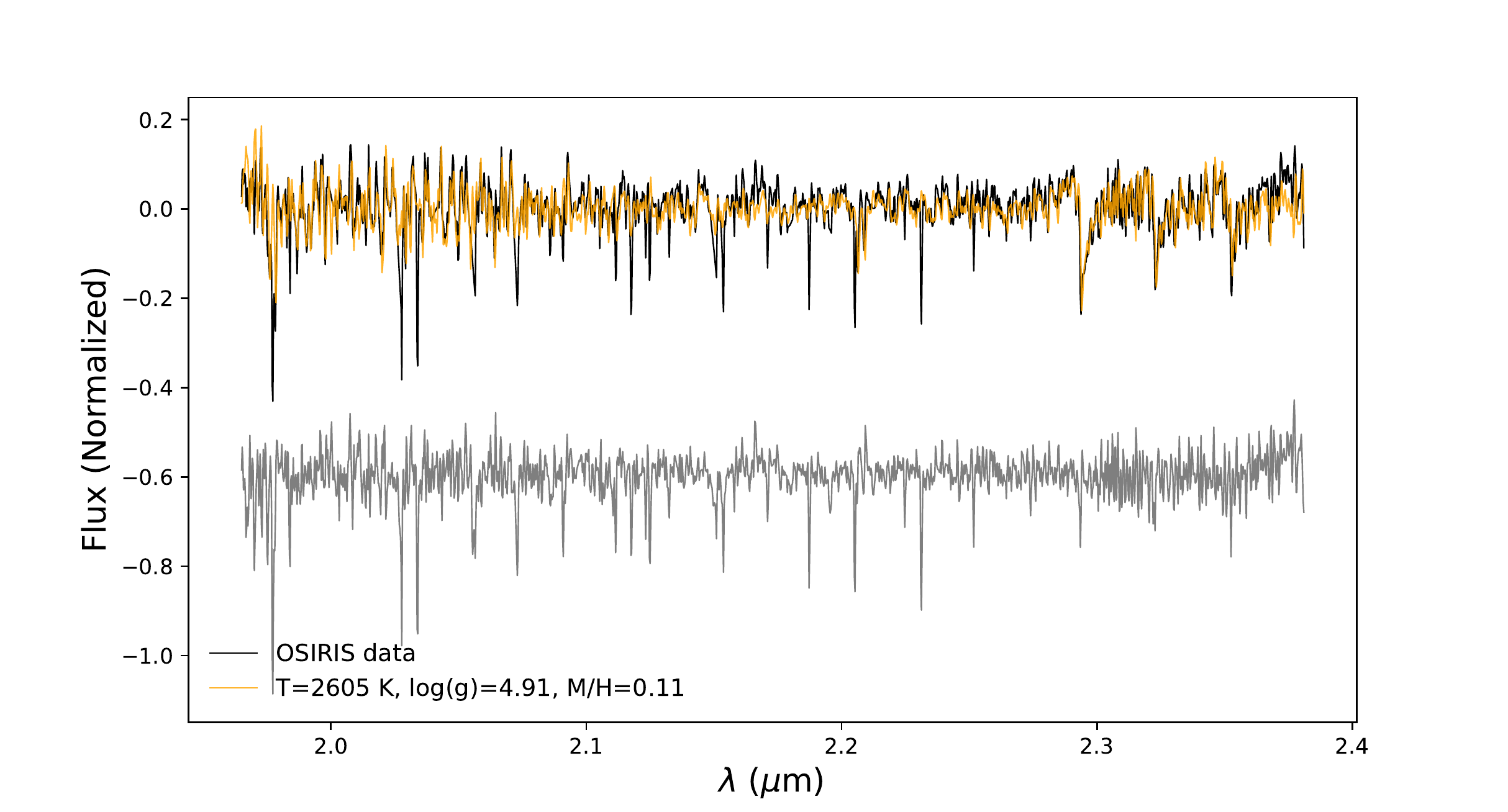}
\caption{Our fully reduced, combined, and continuum subtracted moderate resolution OSIRIS $K$-band spectra of HD 284149 AB b in black plotted against the best-fit model in orange. The residuals are plotted below in gray.}
\label{fig:hd284_sub_fit}
\end{figure*}

\begin{figure*}
\centering
  \includegraphics[width=\textwidth]{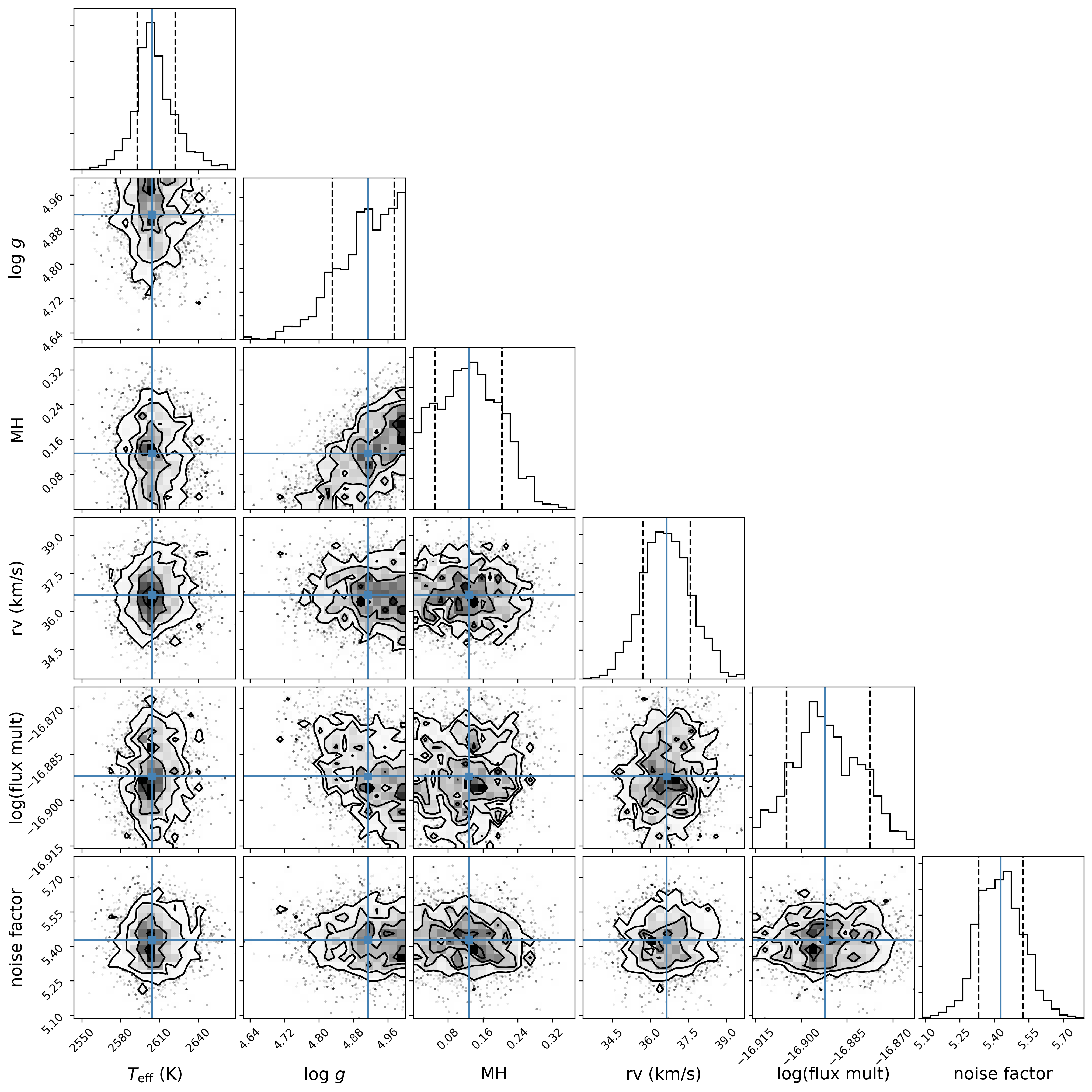}
\caption{Corner plot from our MCMC fits for our continuum subtracted OSIRIS $K$-band spectra.}
\label{fig:hd284_sub_corner}
\end{figure*}

We then ran our fitting procedure on the continuum subtracted data using the same grid. We employed the same filtering code used to flatten the data on the model grid to flatten the models. The best fit parameters from our continuum subtracted $K$-band OSIRIS data are $T_\mathrm{eff}$ = 2605$^{+19}_{-13}$ K, $\log g$ = 4.91$^{+0.060}_{-0.074}$, and [M/H] = 0.11$^{+0.076}_{-0.067}$. Figures \ref{fig:hd284_sub_fit}--\ref{fig:hd284_sub_corner} show the best-fit model for the continuum subtracted data and the corresponding corner plot from our MCMC analysis. Our continuum subtracted results are similar to our continuum results, but are not consistent within the error bars. In order to encompass the results from both fits, we adopt values of $T_\mathrm{eff} = 2502$~K, with a range of 2291--2624~K, $\log g=4.52$, with a range of 4.38--4.91, and [M/H] = 0.37, with a range of 0.10--0.55 summarized in Table \ref{tab:atm_param}.

\begin{deluxetable*}{lccc} 
\tabletypesize{\scriptsize} 
\tablewidth{0pt} 
\tablecaption{Summary of atmospheric parameters derived from MCMC fits.\label{tab:fits}}
\label{tab:atm_param}
\tablehead{ 
  \colhead{Spectra} & \colhead{Effective Temperature} & \colhead{Surface Gravity} &
  \colhead{Metallicity} \\
  \colhead{HD 284149 AB b} & \colhead{${T}_\mathrm{{eff}}$ (K)} & \colhead{$\log g$} & \colhead{[M/H]} } 
\startdata 
\multicolumn{3}{c}{PHOENIX-ACES} \\
\hline
OSIRIS Including Continuum & 2502$^{+19}_{-11}$ & 4.52$^{+0.22}_{-0.14}$  & 0.37$^{+0.18}_{-0.12}$ \\
OSIRIS Continuum Subtracted & 2605$^{+19}_{-13}$ & 4.91$^{+0.060}_{-0.074}$ & 0.11$^{+0.076}_{-0.067}$ \\
\hline
Adopted Values & $2502$ & 4.52 & 0.37 \\
Allowed Range of Values & 2291--2624 & 4.38--4.91 & 0.10--0.55  \\
\enddata

\tablenotetext{-}{Based on our fitting, we have chosen to adopt the best-fit values from the G\"ottingen spectral library fit to the OSIRIS data with the continuum included.  However, we allow for uncertainties that encompass the range of most fits to both the continuum data and the continuum subtracted data.}
\end{deluxetable*}

\subsection{C and O Abundances}
With best fit values for temperature, surface gravity, and metallicity, we can derive abundances of C and O using our OSIRIS $K$-band spectra. We chose the best-fit values from the continuum fits of the data because the the results from either fit were fairly similar, and the continuum contains important information about the temperature when not biased by speckles, as is the case here. We constructed a mini-grid using $PHOENIX$ that holds our derived parameters the same, but varies the abundance of C and O. C and O were each varied from 0 to 1000 times Solar using a uniform logarithmic sampling, resulting in 12 synthetic spectra. We fit for the abundance of C first, holding O at its initial value. Next, the C abundance was set to its nominal value, and we fit for O.  We choose not to vary the bulk atmospheric parameters (temperature, log(g), and metallicity) in this grid because they are constrained by the continuum.  A change in C or O abundance will not change the shape of the continuum to result in an improved fit, but will instead impact line depths.  While the depths of the lines are impacted by bulk parameters, it would not be possible to jointly improve the continuum and the line depths by only varying the C and O abundance.  Indeed, this effect was seen in \citet{konopacky2013}, where a grid with variation in temperature, gravity, and C and O was used but all best fits were at the same temperature and gravity as determined using the continuum information. 

\begin{figure*}
\centering
  \includegraphics[scale=0.65]{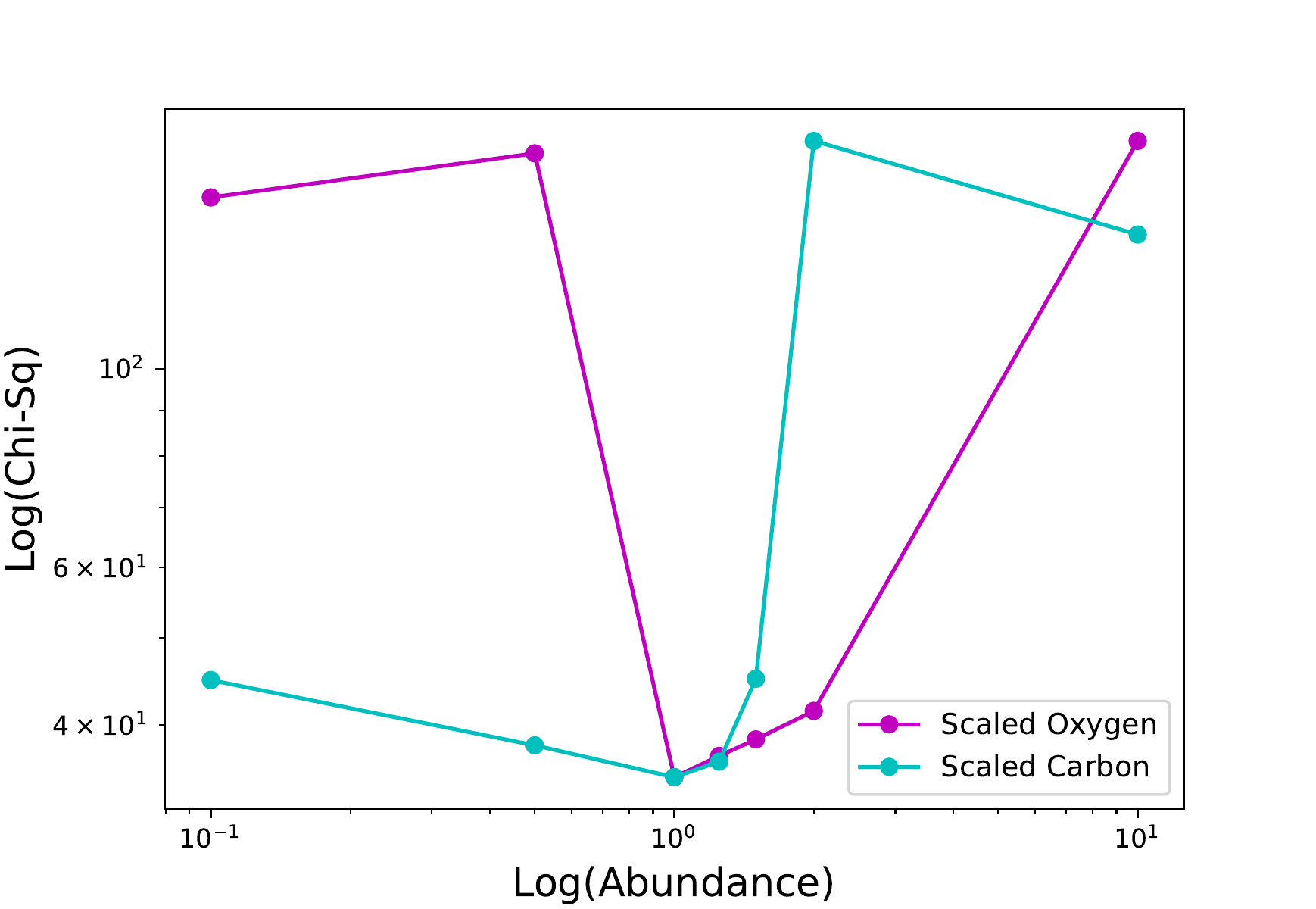}
\caption{Results of $T_\mathrm{eff}$ = 2502$^{+19}_{-11}$ K, $\log g$ = 4.52$^{+0.22}_{-0.14}$, and [M/H] = 0.37$^{+0.18}_{-0.12}$ model fits with varying abundances for both C and O to our continuum-subtracted OSIRIS spectrum. The abundances are given in units relative to the ratio in the Sun, such that a value of zero implies the solar value.  The scalings of C prefer solar as well as the scalings of O prefer solar values. From these fits we find C/O = 0.59$^{+0.15}_{-0.30}$.}
\label{fig:chisq_hd284}
\end{figure*}

Figure \ref{fig:chisq_hd284} shows the resulting \(\chi^2\) distribution as a function of C and O abundance. The models with the lowest \(\chi^2\) when compared to the flattened data gave us the best-fits for both C and O. The best fit for C had an abundance scaling of 1.00$^{+0.30}_{-0.45}$, and the best fit for O had a lower limit abundance scaling of solar or 1.00$^{+0.37}_{-0.03}$.  To calculate the 1-$\sigma$ uncertainties in each mole fraction value, we used the values from models within $\pm$1 of our lowest \(\chi^2\).

\subsection{C/O Ratio for HD 284149 AB b}
In our previous work, we have used OSIRIS data to constrain the C/O ratios of directly imaged companions such as HR 8799 b, c, and d, $\kappa$ Andromedae b and VHS 1256 b (e.g., \citealt{konopacky2013,wilcomb2020,ruffio2021,hoch2022}). For giant planets formed by rapid gravitational instabilities, their atmospheres should have elemental abundances that are similar to their host stars \citep{helled2009}. If giant planets form by a core/pebble accretion process, there could be a range of elemental abundances possible \citep{oberg2011,madhu2019}. In this framework, the abundances of giant planets' atmospheres formed by core/pebble accretion are highly dependent on the location of formation relative to CO, CO$_2$, and H$_2$O frost lines and the amount of solids collected during runaway accretion phase. This can potentially be diagnosed using the C/O ratio. 

The C/O ratio is dependent on the abundances of C and O in the atmosphere. The equation used for this derivation is 


\[\frac{C}{O}=10^{\epsilon_{C}-\epsilon_{O}},\]


\noindent where $\epsilon$ is the scaled abundance relative to Solar. The C/O ratio we derive for HD 284149 AB b is 0.59$^{+0.15}_{-0.30}$.
\subsection{Dynamical Masses of the HD~284149 System} \label{sec:dyn_mass}

Previous calculations of mass values of the HD~284149 system calculated by \cite{bonavita2017} are derived from the measured photometry of the stellar and substellar components. The adopted mass values in \cite{bonavita2017} are $0.16\pm0.04~M_{\odot}$ for HD~284149~B and $26\pm3~M_{\rm{Jup}}$ for HD~284149~AB~b. Photometric mass values are inherently dependent on the stellar age, which can drive large uncertainties, or sometimes even discrepancies, between the derived photometric model-dependent mass and the dynamical model-independent mass. Model-independent dynamical masses of stellar and substellar companions, that do not depend on a stellar age estimate, can be determined from orbital information.  In particular, combining astrometric accelerations from Hippacros \citep{1997A&A...323L..49P} and \emph{Gaia} eDR3 \citep{2021A&A...649A...1G} with measured relative astrometry from direct imaging as demonstrated in  \cite{2022A&A...668A.140R} has been shown to be a powerful way to estimate companion masses. 

We have calculated preliminary dynamical masses of HD~284149~B and HD~284149~AB~b using the astrometric accelerations from the Hipparcos-\emph{Gaia} catalog of accelerations \citep[HGCA;][]{2021ApJS..254...42B} and the relative astrometry from \cite{bonavita2017}. This astrometric information can be combined using \texttt{orvara} \citep{2021AJ....162..186B} which is an orbit-fitting code specifically designed to combine absolute astrometric data from Hipparcos and \emph{Gaia}, with relative astrometry from direct imaging that has the capability of fitting multi-planet systems.

\begin{figure*}
    \centering
    \includegraphics[width=0.48\textwidth]{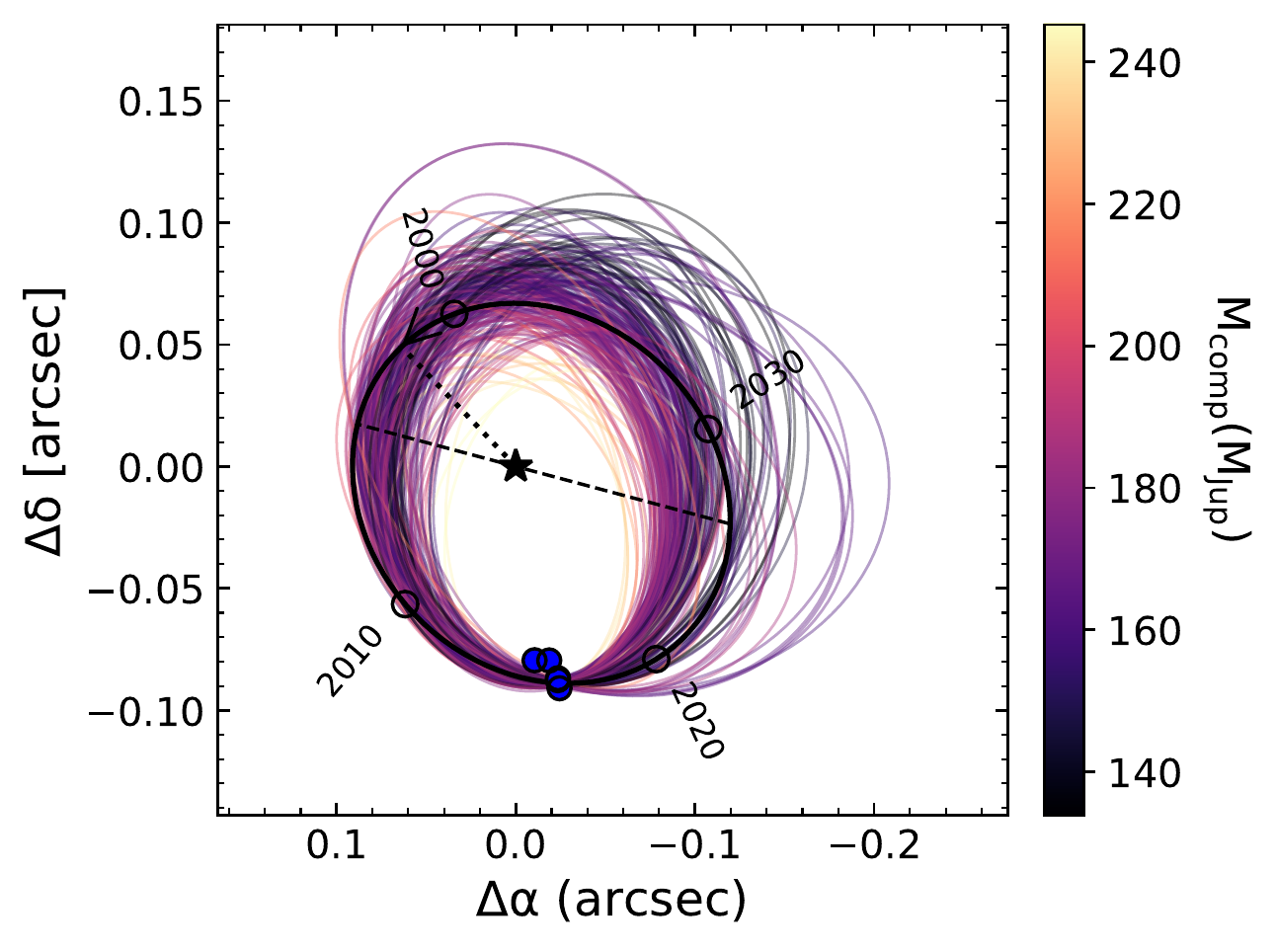}
    \includegraphics[width=0.45\textwidth]{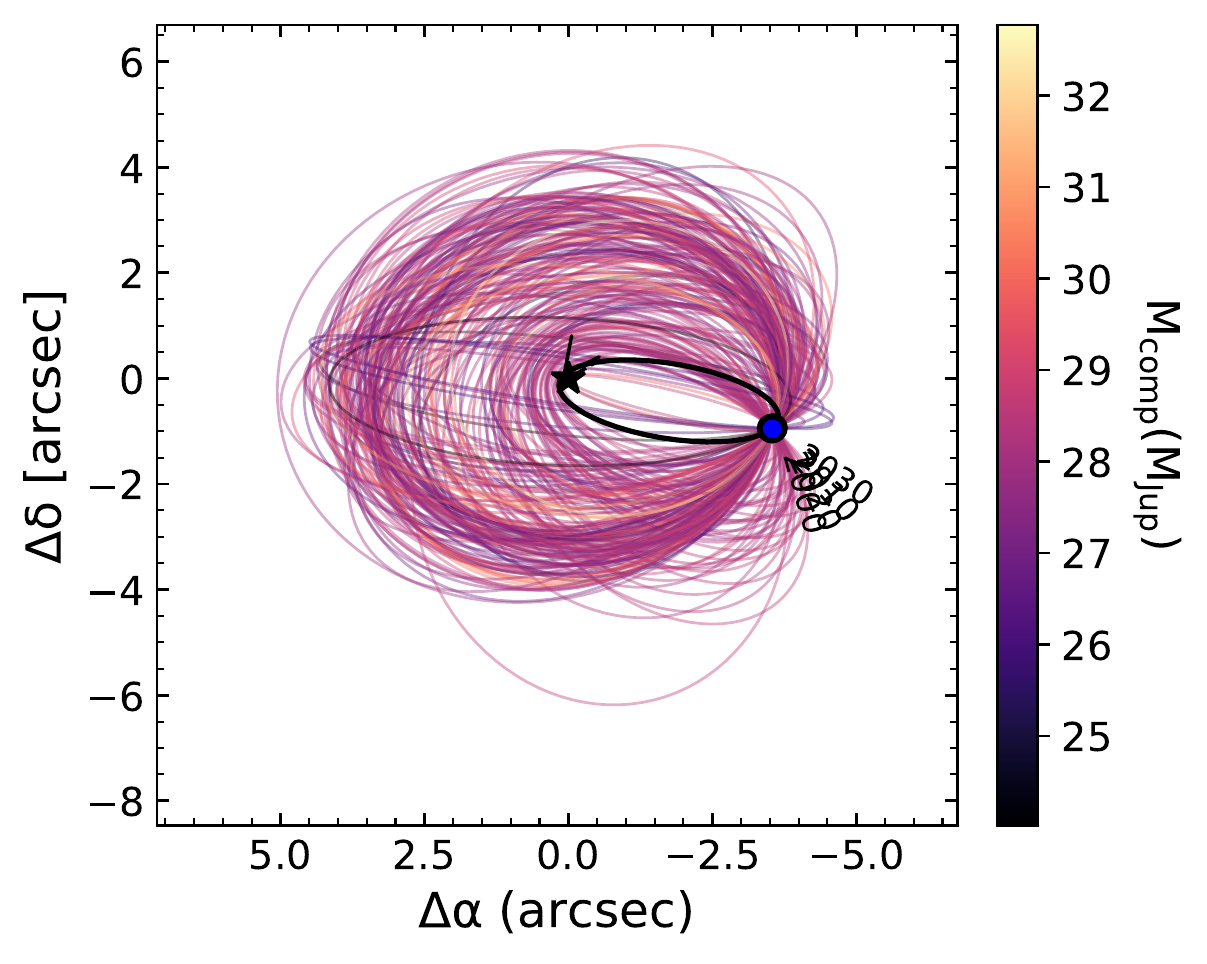}
    \caption{
    The relative astrometric orbit of HD~284149~B (\emph{left}) and HD~284149~AB~b (\emph{right}) The thick black line represents the highest likelihood orbit; the thin colored lines represent 200 orbits drawn randomly from the posterior distribution. Dark purple corresponds to a low companion mass and light yellow corresponds to a high companion mass. The dotted black line shows the periastron passage and the arrow at the periastron passage shows the direction of the orbit. The dashed line indicates the line of nodes. Predicted past and future relative astrometric points are shown by black circles with their respective years, while the observed relative astrometric points from VLT/SPHERE data \citep{bonavita2017} are shown by the blue-filled data point, where the measurement error is smaller than the plotted symbol.
    }
    \label{fig:orbit_plots}
\end{figure*}

\begin{figure*}
    \includegraphics[width=0.95\textwidth]{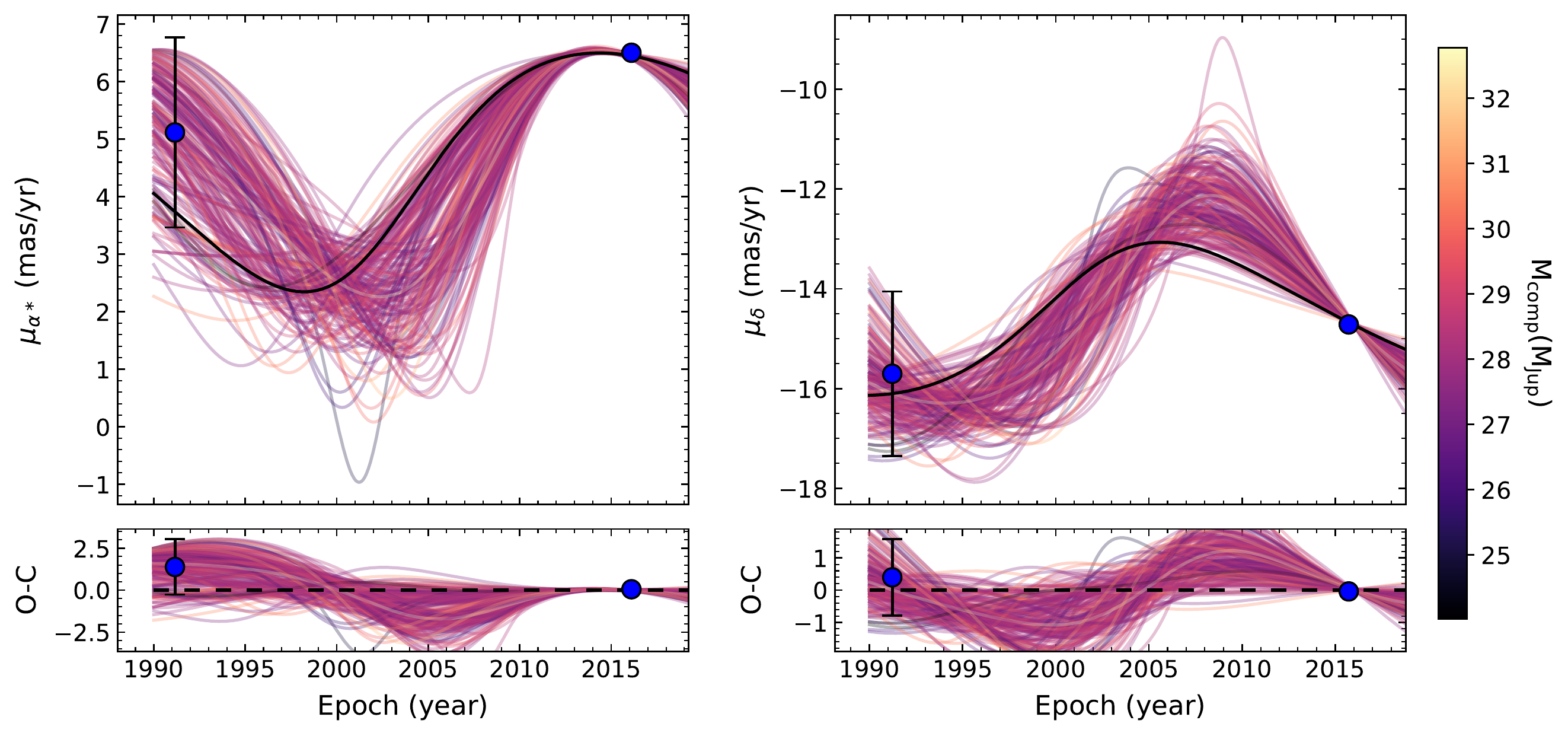}
    \caption{Acceleration induced by the companion on the host star as measured from absolute astrometry from Hipparcos and \emph{Gaia}. The thick black line represents the highest likelihood orbit; the thin colored lines are 200 orbits drawn randomly from the posterior distribution. Darker purple represents a lower companion mass and light yellow represents a higher companion mass for HD~284149~AB~b. The residuals of the proper motions are shown in the bottom panels. 
    }
    \label{fig:astrometric_acc}
\end{figure*}

\begin{figure*}
    \includegraphics[width=\textwidth]{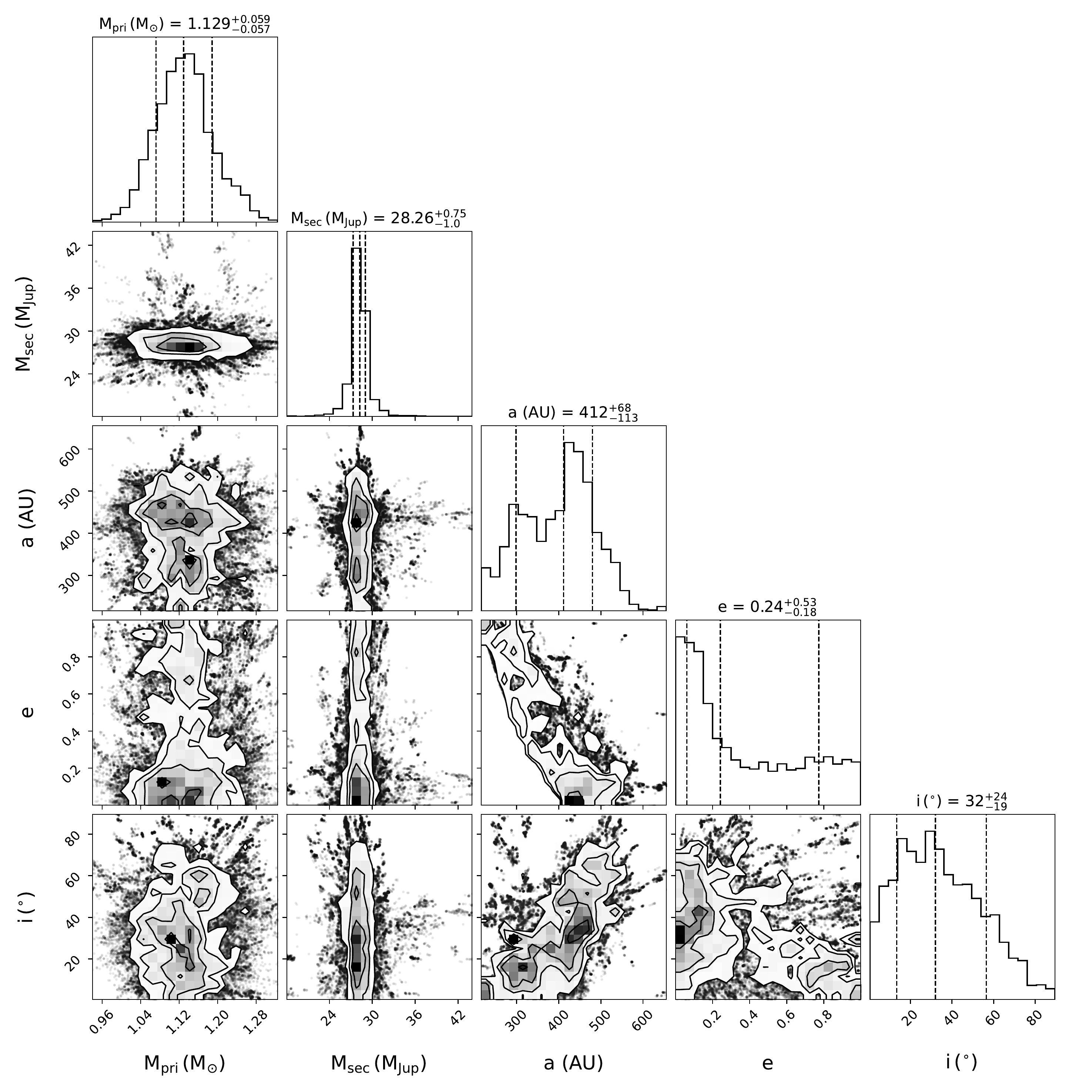}
    \caption{Corner plot showing the fitted orbital parameters for HD~284149~AB~b using the orbit-fitting code \texttt{orvara} \citep{2021AJ....162..186B}.}
    \label{fig:corner_HD284149}
\end{figure*}

\begin{figure*}
    \includegraphics[width=\textwidth]{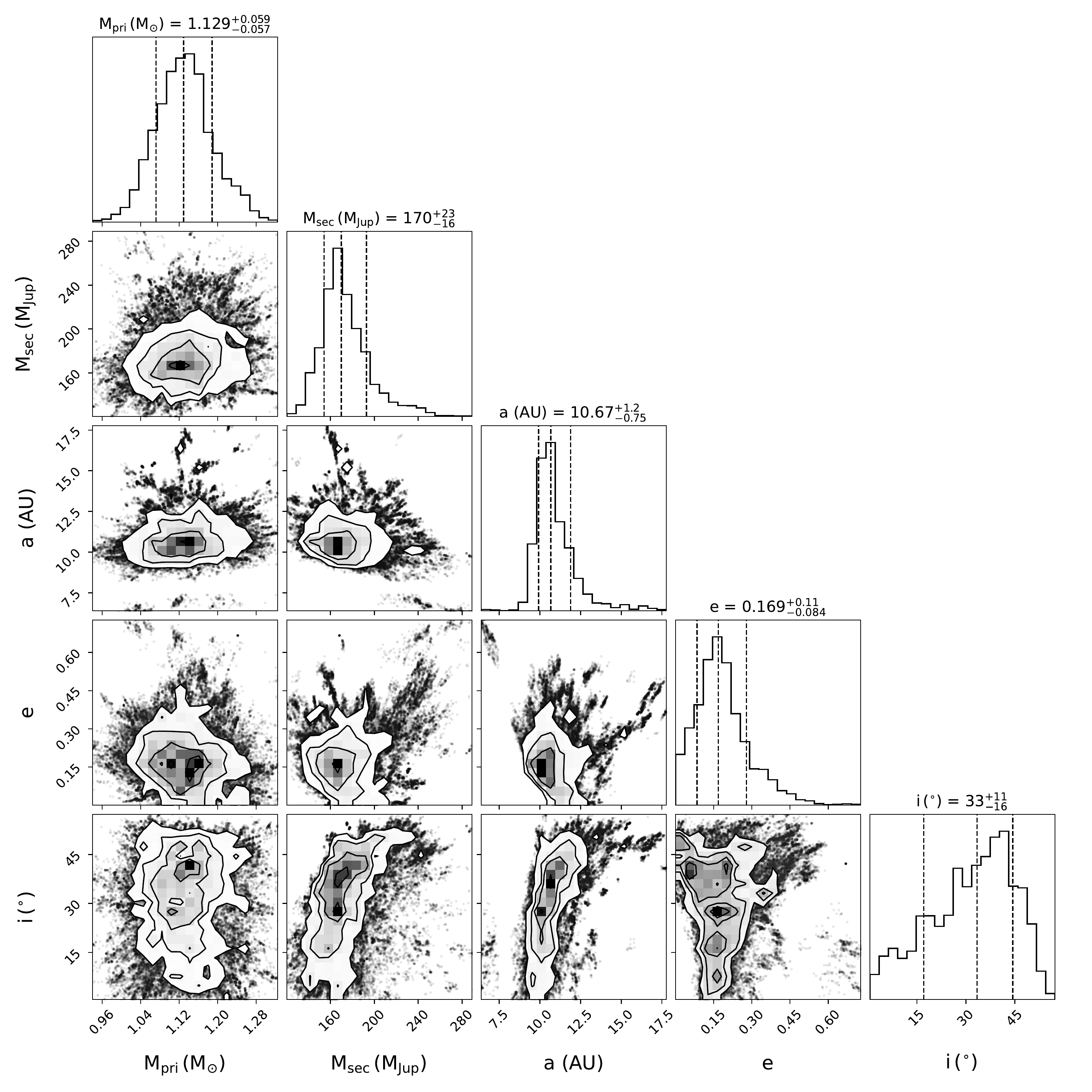}
    \caption{Corner plot showing the fitted orbital parameters for HD~284149~B using the orbit-fitting code \texttt{orvara} \citep{2021AJ....162..186B}.}
    \label{fig:corner_HD284149_BD}
\end{figure*}

We ran the MCMC orbital fit with 500,000 steps in each chain for both HD~284149~AB~b and HD~284149~B, with a prior set on the mass of the primary star HD~284149~A of $1.14\pm0.05~M_{\odot}$ \citep{bonavita2014}. The resulting orbital plots are shown in Figure \ref{fig:orbit_plots} and the fits to the astrometric accelerations are shown in Figure \ref{fig:astrometric_acc}, with the corner plots for HD~284149~B and HD~284149~AB~b are shown in Figure \ref{fig:corner_HD284149} and Figure \ref{fig:corner_HD284149_BD} respectively.

Using this approach we derive a dynamical mass of the inner companion HD~284149~B of $M = 170^{+23}_{-16}~M_{\textrm{Jup}}$ (or $0.16\pm0.02~M_{\odot}$) which is in agreement with the photometric value derived in \cite{bonavita2017} of $0.16\pm0.04~M_{\odot}$. We also derive a dynamical mass of the substellar companion HD~284149~AB~b of $M= 28.26^{+0.75}_{-1.00}~M_{\textrm{Jup}}$, which is also in agreement with the photometrically-derived mass of $26\pm3~M_{\rm{Jup}}$ from \cite{bonavita2017}.  The close correspondence of the dynamical mass estimate and the photometrically-derived mass is encouraging, and provides additional weight to the estimated age of the system.

We note that in spite of the small error bars on the derived mass, there might be biases in this result due to low phase coverage of astrometry for this orbit.  Even with the long time baseline of Hipparcos and Gaia, the estimated orbital period of $\sim$7300 years for the companion means that the orbit phase coverage is only 0.3\%.  Minimal orbit coverage can result in biases in the resulting orbital parameters (e.g., \citealt{Kosmo2019,Ferrer2021,DoO2023}), particularly when there are systematics in the astrometric data that are not accounted for.  It is outside the scope of this paper to explore the orbital properties and astrometry of this sytem in detail, but we take the derivation of a similar dynamical mass to the photometric mass as an indication that using the current system age estimate is reasonable. Therefore, we use the photometrically derived mass of HD~284149~AB~b in our C/O ratio analysis in Section \ref{sec:c-o_analysis}.

\subsection{Non-Detections}\label{nondetections}
For our imaging spectroscopy survey of exoplanetary atmospheres with Keck/OSIRIS we have thus far gathered $K$-band data on eight directly imaged planets; $\kappa$ And b, VHS 1256 b, HR 8799 bcd, HD 284149 AB b, GJ 504 b, and 51 Eri b. GJ 504 b and 51 Eri b were not detected in our OSIRIS data and details of the observations are in Table \ref{tab:log-nondetect}. GJ 504 b is a Jovian planet of 4$^{+4.5}_{-1.0}\,M_{\mathrm{Jup}}$ orbiting at a projected separation of 43.5 au around the Sun-like G0-type star, GJ 504 \citep{kuzuhara2013}. GJ 504 b is significantly cooler than other imaged exoplanets with an effective temperature of about 510$^{+30}_{-20}$ K. GJ 504 b is also the first directly imaged planet around a metal-rich host star. 51 Eri b is a gas giant planet that was discovered orbiting the $\sim$20 Myr 51 Eridani star at 13 au \citep{macintosh2015}. 51 Eri b is about 600-750 K with a mass of about 2--12\,$M_{\mathrm{Jup}}$. Both of these planets are lower temperature and lower mass than the majority of the directly imaged companions, and would be excellent candidates for atmospheric characterization to compare to their hotter counterparts.  

For GJ 504 b, we conducted observations in the $K$ broadband mode (1.965--2.381 $\mu$m) with a spatial sampling of 0.050$\arcsec$ per lenslet for this object with the hope of enhancing its detectability via coarser spatial sampling. We observed a telluric calibrator star (HIP 65599) and obtained sky frames in close time to observations of the object. Data cubes (x, y, $\lambda$) were generated after operating the OSIRIS DRP using the observatory provided rectification matrices for the same time frame of the observations, following the steps described in section 4.2. After reduction and telluric correction, we tried to locate the planet in the data cubes.  The separation of GJ 504b means that the speckles were not as bright as some of our other targets, so our attempt to locate the planet was primarily without any speckle removal. We did not see a clear signal when looking at the datacubes. We then tried a cross-correlation approach using a \textit{PHOENIX} model of similar atmospheric parameters (\cite{ruffio2019}) and still did not see a clear signal in our OSIRIS data. 

For 51 Eri b, we conducted observations in the $K$ broadband mode (1.965--2.381 $\mu$m) with a spatial sampling of 0.020$\arcsec$ per lenslet, observed a telluric standard star (HIP 25453) and sky frames in close time to object observations. Data cubes were generated using the OSIRIS DRP as with the other sources. We did not see a clear signal in the raw data, and we also tried a cross-correlation approach using a \textit{PHOENIX} model of similar atmospheric parameters and did not detect the planet.  Given that the detections of these targets is currently beyond what we can achieve with OSIRIS, both GJ 504 b and 51 Eri b would be great candidates for JWST and upcoming extremely large telescopes (ELTs) that may have better sensitivity to objects of high contrast and/or low temperature. 

\begin{deluxetable*}{lcccccc} 
\tabletypesize{\scriptsize} 
\addtolength{\tabcolsep}{0pt}
\tablewidth{0pt} 
\tablecaption{Log of two non-detected planets observed for the Imaging Spectroscopy Survey of Exoplanetary Atmospheres with Keck/OSIRIS.
\label{tab:log-nondetect} }
\tablehead{ 
  \colhead{Companion} & \colhead{Band} & \colhead{Spatial Sampling} & \colhead{Int. Time per} & \colhead{Total Number} & \colhead{Total Int. Time} & \colhead{Dates Taken} \\
 \colhead{} & \colhead{} & \colhead{(mas)} & \colhead{Frame (sec)} & \colhead{of Frames} & \colhead{(mins)} & \colhead{}
}
\startdata 
51 Eri b & $K$ & 0.02 & 600 & 68 & 680 & Nov 6-8, 2016 $\&$ Nov 3-4 2017  \\
GJ 504 b & $K$ & 0.05 & 900 & 18 & 435 & May 14, 2017 $\&$ May 26-27, 2019 \\
\enddata
\end{deluxetable*}

\section{C/O Ratio Analysis}\label{sec:c-o_analysis}

\begin{deluxetable*}{lcccccc} 
\tabletypesize{\scriptsize} 
\addtolength{\tabcolsep}{0pt}
\tablewidth{0pt} 
\tablecaption{Directly Imaged and Transit/Eclipse System and Atmospheric Parameters }
\tablehead{ 
  \colhead{Companion} & \colhead{Separation (au)} & \colhead{Age (Myr)} & \colhead{Companion Mass (M$_{\mathrm{Jup}}$)} & \colhead{Host Mass (M$_\odot$)} & \colhead{C/O} & \colhead{References}\\
}
\startdata 
\textbf{Directly Imaged Planets} & & & & & & \\
$\kappa$ And b & 55$\pm$2 & 47$\pm$40 & 20$\pm$10 & 2.8$\pm$0.4 & 0.704$^{+0.09}_{-0.24}$ & \tablenotemark{1}, \tablenotemark{2}, \tablenotemark{3}, \tablenotemark{4}\\
VHS 1256 b & 180$\pm$9 & 200$\pm$100 & 12$\pm$6 & 0.152$\pm$0.01 & 0.590$^{+0.28}_{-0.24}$ & \tablenotemark{5}, \tablenotemark{6}, \tablenotemark{7}\\
HD 284149 AB b & 431$\pm$7 & 35$\pm$10 & $26 \pm 3$ & 1.13$\pm$0.06 & 0.589$^{+0.148}_{-0.295}$ & \tablenotemark{8}, \tablenotemark{9} \\
HR 8799 b & 68$\pm$2 & 40$\pm$5 & 5.84 $\pm 0.3$  & 1.47$^{+0.12}_{-0.17}$ & 0.578$^{+0.004}_{-0.005}$ & \tablenotemark{10}, \tablenotemark{64}, \tablenotemark{11}, \tablenotemark{12}, \tablenotemark{13}, \tablenotemark{65}, \tablenotemark{66}\\
HR 8799 c & 42.81$\pm$1.16 & 40$\pm$5 & 7.63$^{+0.64}_{-0.63}$ & 1.47$^{+0.12}_{-0.17}$ & 0.562$\pm$0.004 & \tablenotemark{10}, \tablenotemark{11}, \tablenotemark{12}, \tablenotemark{13},\tablenotemark{65}, \tablenotemark{66} \\
HR 8799 d & 26.97$\pm$0.73 & 40$\pm$5 & 9.81$\pm 0.08$ & 1.47$^{+0.12}_{-0.17}$ & 0.551$^{+0.005}_{-0.004}$ & \tablenotemark{10}, \tablenotemark{11}, \tablenotemark{12}, \tablenotemark{13}, \tablenotemark{65}, \tablenotemark{66}\\
HR 8799 e & 17$\pm$0.5 & 40$\pm$5 & 7.64$^{+0.89}_{-0.91}$ & 1.47$^{+0.12}_{-0.17}$ & 0.60$^{+0.07}_{-0.08}$ & \tablenotemark{10}, \tablenotemark{11}, \tablenotemark{12}, \tablenotemark{14}, \tablenotemark{65}, \tablenotemark{66}\\
HIP 65426 b & 92.3$\pm$0.2 & 14$\pm$3 & 7$\pm$2 & 1.96$\pm$0.04 & 0.55$^{+0}_{-0.55}$ & \tablenotemark{16},\tablenotemark{15} \\
TYC 8998-760-1 b & 162$\pm$0.28 & 16.7$\pm$1.4 & 14$\pm$3 & 1.00$\pm$0.02 & 0.52$^{+0.04}_{-0.03}$ & \tablenotemark{17}, \tablenotemark{18} \\
AB Pic b & 273$\pm$2 & 13.3$\pm$1.1 & 10$\pm$1 & 0.84$\pm$0.11 & 0.58$\pm0.08$ & \tablenotemark{21}, \tablenotemark{22}, \tablenotemark{23}, \tablenotemark{19}\\
\hline
\textbf{Transiting Planets} & & & & & & \\
CoRot-1 b & 0.0261$\pm$0.0005 & 1600$\pm$500 & 1.13$\pm$0.07 & 1.22$\pm$0.03 & 0.9$^{+0.7}_{-0.3}$ & \tablenotemark{24}, \tablenotemark{25}, \tablenotemark{26}, \tablenotemark{27}\\
HAT-P-2 b & 0.06814$\pm$0.00051 & 1440$\pm$470 & 8.7$\pm$0.2 & 1.33$\pm$0.03 & 0.5$^{+0.3}_{-0.2}$ & \tablenotemark{28}, \tablenotemark{26}, \tablenotemark{27}\\
HAT-P-7 b & 0.03813$\pm$0.00036 & 2070$\pm$300 & 1.84$\pm$0.53 & 1.51$\pm$0.04 & 1.4$^{+0.3}_{-0.3}$ & \tablenotemark{29}, \tablenotemark{26}, \tablenotemark{27} \\
HAT-P-32 b & 0.03397$\pm$0.00051 & 2700$\pm$800 & 0.68$\pm$0.1 & 1.132$\pm$0.05 & 1.5$^{+0.3}_{-0.3}$ & \tablenotemark{30}, \tablenotemark{31}, \tablenotemark{27}\\
HAT-P-41 b & 0.04258$\pm$0.00047 & 2200$\pm$400 & 0.795$\pm$0.056 & 1.418$\pm$0.047 & 1.6$^{+0.3}_{-0.3}$ & \tablenotemark{32}, \tablenotemark{26}, \tablenotemark{27}\\
HAT-P-70 b & 0.04739$\pm$0.00031 & 600$\pm$30 & 6.78$\pm$0 & 1.89$\pm$0.01 & 1$^{+0.08}_{-0.08}$ & \tablenotemark{33}, \tablenotemark{27}\\
HD 189733 b & 0.03126$\pm$0.00036 & 6800$\pm$500 & 1.13$\pm$0.08 & 0.812$\pm$0.04 & 0.8$^{+0.1}_{-0.2}$ & \tablenotemark{34}, \tablenotemark{35}, \tablenotemark{36}, \tablenotemark{27}\\
HD 209458 b & 0.04634$\pm$0.0007 & 3100$\pm$800 & 0.73$\pm$0.04 & 1.19$\pm$0.28 & 0.92$^{+0.01}_{-0.01}$ & \tablenotemark{37}, \tablenotemark{26}, \tablenotemark{38}, \tablenotemark{27}\\
KELT-1 b & 0.02466$\pm$0.00016 & n/a & 27.23$\pm$0.5 & 1.3766$\pm$0.25 & 0.4$^{+0.2}_{-0.2}$ & \tablenotemark{39}, \tablenotemark{40}, \tablenotemark{27} \\
KELT-7 b & 0.04415$\pm$0.0006 & 1300$\pm$200 & 1.39$\pm$0.22 & 1.45$\pm$0.05 & 1.6$^{+0.3}_{-0.3}$ & \tablenotemark{41}, \tablenotemark{35}, \tablenotemark{40}, \tablenotemark{27} \\
KELT-9 b & 0.03462$\pm$0.001 & n/a & 2.44$\pm$0.7 & 2.431$\pm$0.3 & 1.1$^{+0.6}_{-0.6}$ & \tablenotemark{42}, \tablenotemark{40}, \tablenotemark{27} \\
Kepler-13A b & 0.03641$\pm$0.00087 & 1120$\pm$100 & 9.28$\pm$1.6 & 1.67$\pm$0.08 & 0.4$^{+0.2}_{-0.2}$ & \tablenotemark{43}, \tablenotemark{44}, \tablenotemark{45}, \tablenotemark{27} \\
TrES-3 b & 0.02282$\pm$0.0003 & 1000$\pm$500 & 1.91$\pm$0.08 & 0.820181$\pm$0.045 & 1.6$^{+0.3}_{-0.3}$ & \tablenotemark{46}, \tablenotemark{26}, \tablenotemark{40}, \tablenotemark{27}\\
WASP-4 b & 0.0226$\pm$0.0007 & 7000$\pm$2900 & 1.186$\pm$0.09 & 0.97$\pm$0.150455 & 1.5$^{+0.3}_{-0.4}$ & \tablenotemark{47}, \tablenotemark{48}, \tablenotemark{27}\\
WASP-12 b & 0.0232$\pm$0.00064 & 2000$\pm$700 & 1.465$\pm$0.079 & 1.17$\pm$0.183504 & 1$^{+0.05}_{-0.05}$ & \tablenotemark{49}, \tablenotemark{50}, \tablenotemark{40}, \tablenotemark{27}\\
WASP-18 b & 0.02024$\pm$0.0003 & 1570$\pm$1000 & 10.2$\pm$0.35 & 1.294$\pm$0.06 & 0.9$^{+0.05}_{-0.05}$ & \tablenotemark{51}, \tablenotemark{52}, \tablenotemark{27}\\
WASP-19 b & 0.01652$\pm$0.0005 & 6400$\pm$4000 & 1.154$\pm$0.08 & 0.965$\pm$0.091 & 0.5$^{+0.2}_{-0.2}$ & \tablenotemark{53}, \tablenotemark{52}, \tablenotemark{27}\\
WASP-33 b & 0.0239$\pm$0.00063 & n/a & 2.093$\pm$0.139 & 1.653$\pm$0.349901 & 0.9$^{+0.05}_{-0.05}$ & \tablenotemark{54}, \tablenotemark{50}, \tablenotemark{27}\\
WASP-43 b & 0.01528$\pm$0.00018 & 7000$\pm$7000 & 2.05$\pm$0.05 & 0.65$\pm$0.105 & 0.7$^{+0.1}_{-0.2}$ & \tablenotemark{55}, \tablenotemark{26}, \tablenotemark{27} \\
WASP-74 b & 0.3443$\pm$0.00036 & 4200$\pm$1600 & 0.826$\pm$0.21 & 1.044$\pm$0.152063 & 1.4$^{+0.4}_{-0.3}$ & \tablenotemark{56}, \tablenotemark{57}, \tablenotemark{27}\\
WASP-76 b & 0.033$\pm$0.0005 & 5300$\pm$6000 & 0.92$\pm$0.03 & 1.226$\pm$0.2333546 & 1.2$^{+0.4}_{-0.1}$ & \tablenotemark{58}, \tablenotemark{40}\\
WASP-77 A b & 0.02335$\pm$0.00045 & 6200$\pm$4000 & 1.667$\pm$0.06 & 0.91$\pm$0.025 & 0.8$^{+0.1}_{-0.1}$ & \tablenotemark{59}, \tablenotemark{52}, \tablenotemark{27}\\
WASP-79 b & 0.519$\pm$0.0008 & 1920$\pm$350 & 0.85$\pm$0.18 & 1.54$\pm$0.36 & 0.3$^{+0.2}_{-0.2}$ & \tablenotemark{60}, \tablenotemark{40}, \tablenotemark{26}, \tablenotemark{27}\\
WASP-103 b & 0.01987$\pm$0.00021 & 4000$\pm$1000 & 1.455$\pm$0.09 & 1.22$\pm$0.039 & 1$^{+0.05}_{-0.05}$ & \tablenotemark{61}, \tablenotemark{26}, \tablenotemark{27}\\
WASP-121 b & 0.02596$\pm$0.00063 & 1500$\pm$1000 & 1.157$\pm$0.07 & 1.368$\pm$0.084 & 1.3$^{+0.1}_{-0.1}$ & \tablenotemark{62}, \tablenotemark{63}, \tablenotemark{27}\\
\enddata
\label{tab:megatable}
\tablerefs{(1) \cite{carson2013}, (2)  \cite{jones2016}, (3)  \cite{hinkley2013}, (4) \cite{wilcomb2020}, (5) \cite{gauza2015},  (6) \cite{dupuy2022}, (7) \cite{hoch2022}, (8) \cite{bonavita2014}, (9) \cite{bonavita2017}, (10) \cite{marois2008}, (11) \cite{zuckerman2011}, (12) \cite{konopacky2013}, (13) \cite{ruffio2021}, (14) \cite{molliere2020}, (15) \cite{petrus2021}, (16) \cite{chauvin2017}, (17) \cite{bohn2020a}, (18) \cite{zhang2021}, (19) \cite{palma-bifani2022}, (20) \cite{booth2021}, (21) \cite{chauvin2005}, (22) \cite{bonnefoy2010}, (23) \cite{bonnefoy2014b}, (24) \cite{barge2008}, (25) \cite{pont2010}, (26) \cite{bonomo2017}, (27) \cite{changeat2022}, (28) \cite{bakos2007}, (29) \cite{pal2008}, (30) \cite{hartman2011}, (31) \cite{wang2019}, (32) \cite{hartman2012}, (33) \cite{zhou2019}, (34) \cite{bouchy2005}, (35) \cite{stassun2017}, (36) \cite{addison2019}, (37) \cite{henry2000}, (38) \cite{rosenthal2021}, (39) \cite{siverd2012}, (40) \cite{TESS}, (41) \cite{bieryla2015}, (42) \cite{gaudi2017}, (43) \cite{borucki2011}, (44) \cite{esteves2015}, (45) \cite{morton2016}, (46) \cite{odonovan2007}, (47) \cite{wilson2008}, (48) \cite{bouma2019}, (49) \cite{hebb2009}, (50) \cite{chakrabartysengupta2019}, (51) \cite{hellier2009}, (52) \cite{cortes-zuleta2020}, (53) \cite{hebb2010}, (54) \cite{colliercameron2010}, (55) \cite{hellier2011}, (56) \cite{hellier2015}, (57) \cite{mancini2019}, (58) \cite{west2016}, (59) \cite{maxted2013}, (60) \cite{smalley2012}, (61) \cite{gillon2014}, (62) \cite{delrez2016}, (63) \cite{bourrier2020}, (64) \cite{bell2015}, (65) \cite{zurlo2022}, (sss) \cite{sepulveda2022}
}
\end{deluxetable*}

\begin{figure*}
\centering
  \includegraphics[scale=0.80]{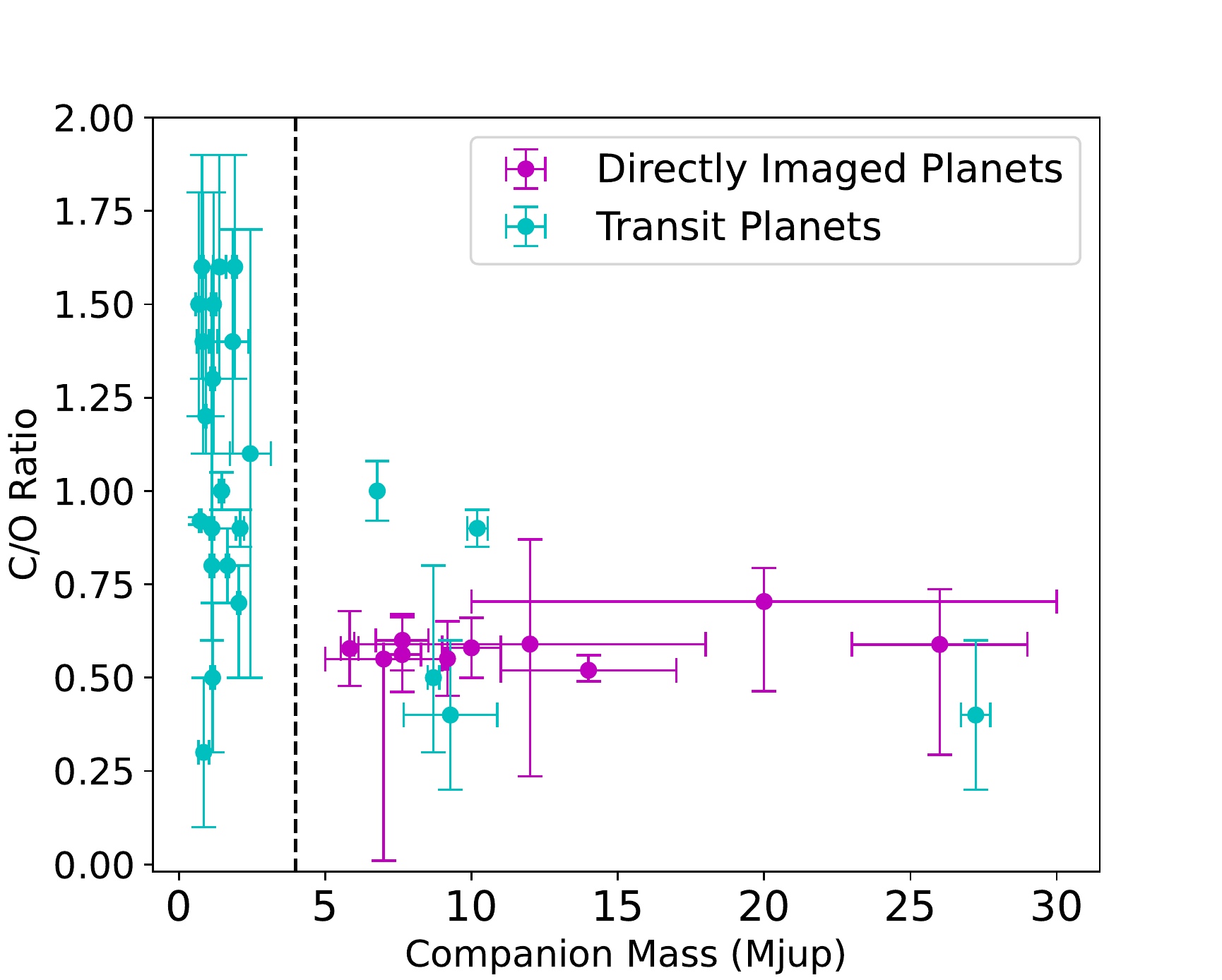}
\caption{Population of 25 transiting exoplanets \citep{changeat2022} in blue and 9 directly imaged planets in magenta. Shown here are their C/O ratios plotted against the companion masses in M$_{\mathrm{Jup}}$. There appear to be two groupings of planets, one with masses below 4 M$_{\mathrm{Jup}}$ and one with masses greater than or equal to 4 M$_{\mathrm{Jup}}$, which is shown by the vertical dashed black line.}
\label{fig:comp_mass_c/o}
\end{figure*}

\begin{figure*}
\centering
  \includegraphics[scale=0.80]{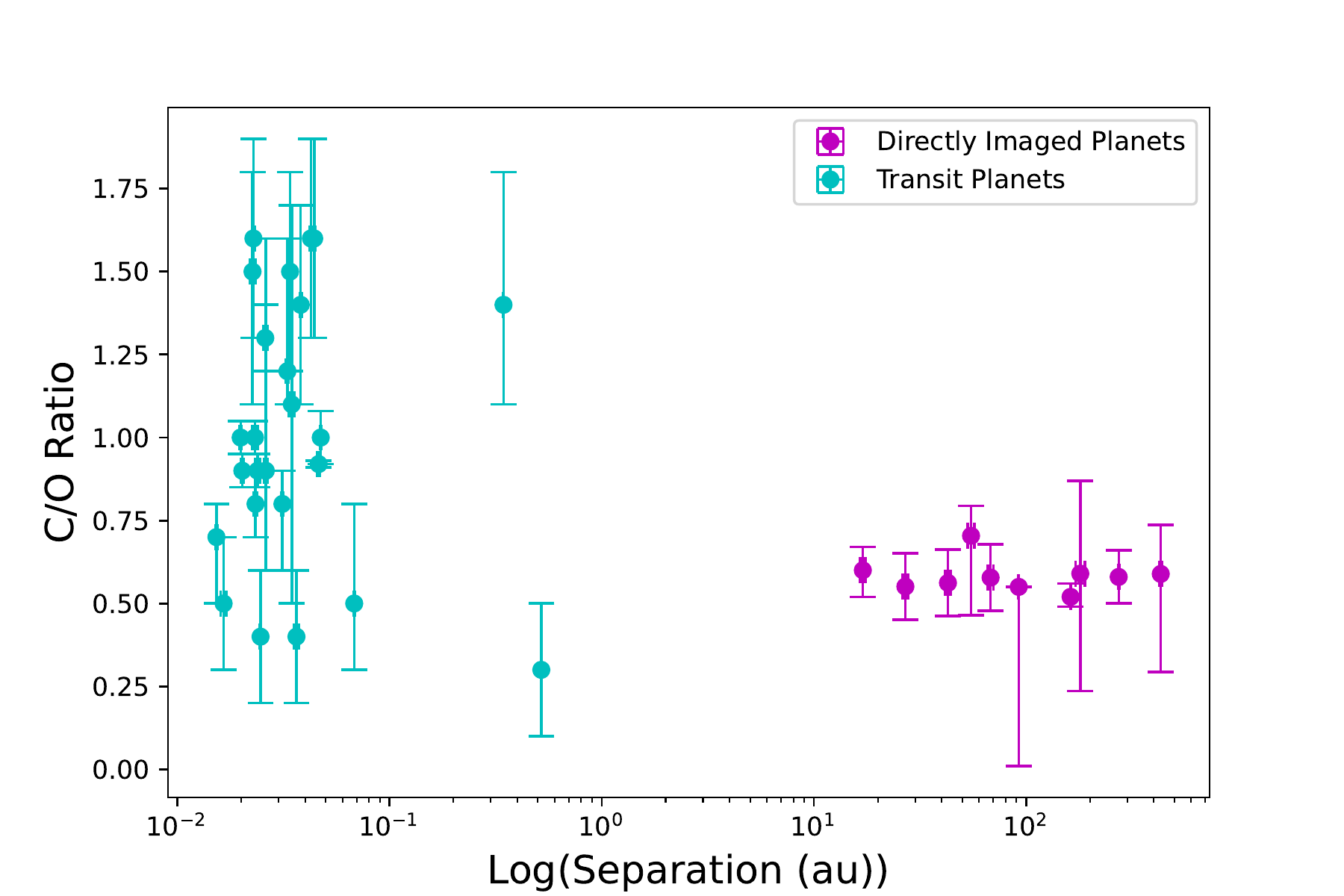}
\caption{Population of 25 retrieval exoplanets in blue and 9 directly imaged planets in magenta. Here are their C/O ratios plotted against their projected separation in au from their host star/system. This plot illustrates highlights the sensitivity to different regions of separation space in each detection methodology for transiting and directly imaged planets.  It also highlights how all directly imaged planets thus far have very consistent C/O ratios, which is not the case for the transiting planets.}
\label{fig:comp_sep_c/o}
\end{figure*}

\begin{figure*}
\centering
  \includegraphics[scale=0.80]{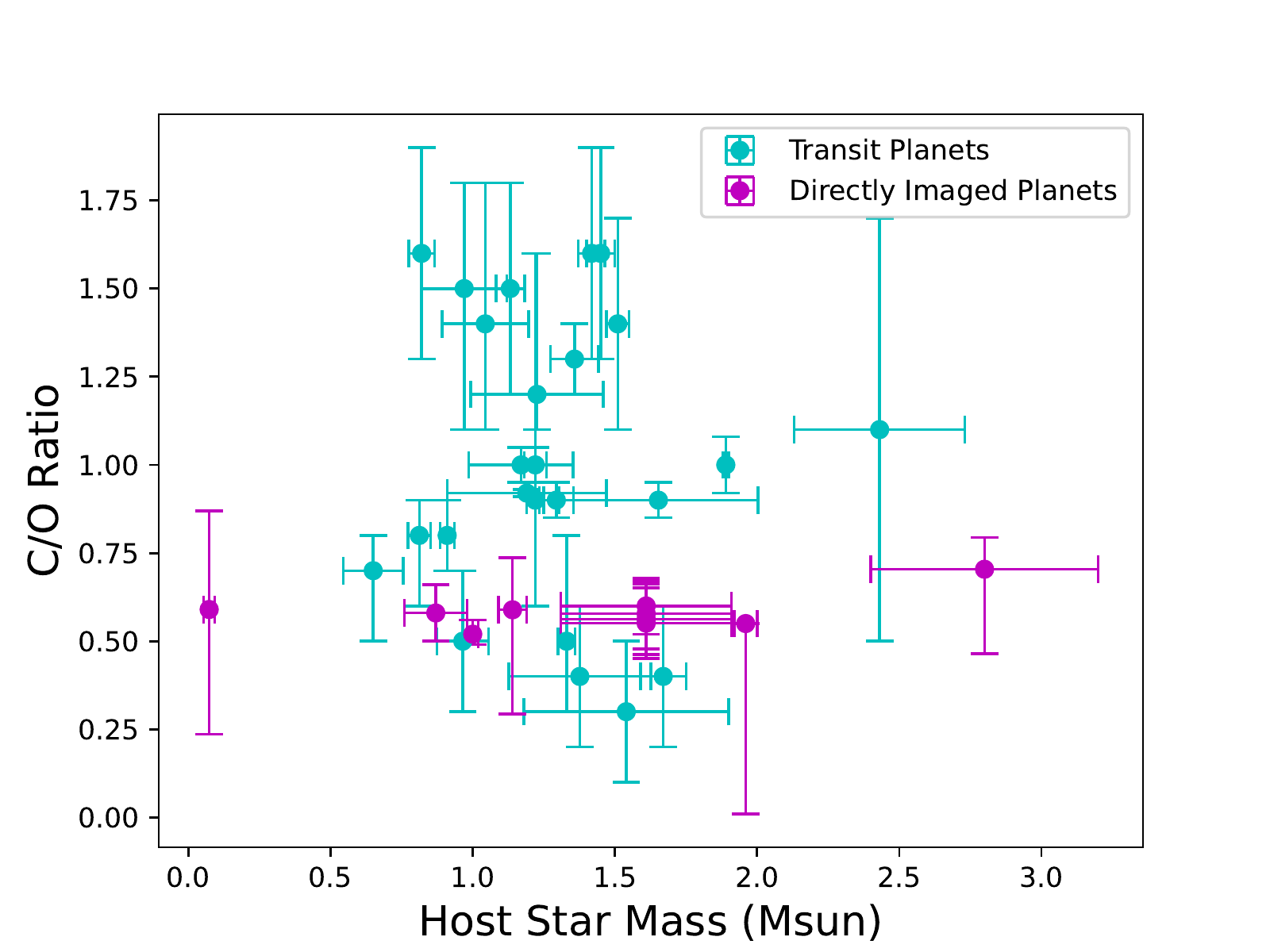}
\caption{Population of 25 retrieval exoplanets in blue and 9 directly imaged planets in magenta. Here are their C/O ratios plotted against the host star masses in M$_{\odot}$. There seems to be no visible correlation between these parameters.}
\label{fig:comp_host_c/o}
\end{figure*}

\begin{figure*}
\centering
  \includegraphics[scale=0.80]{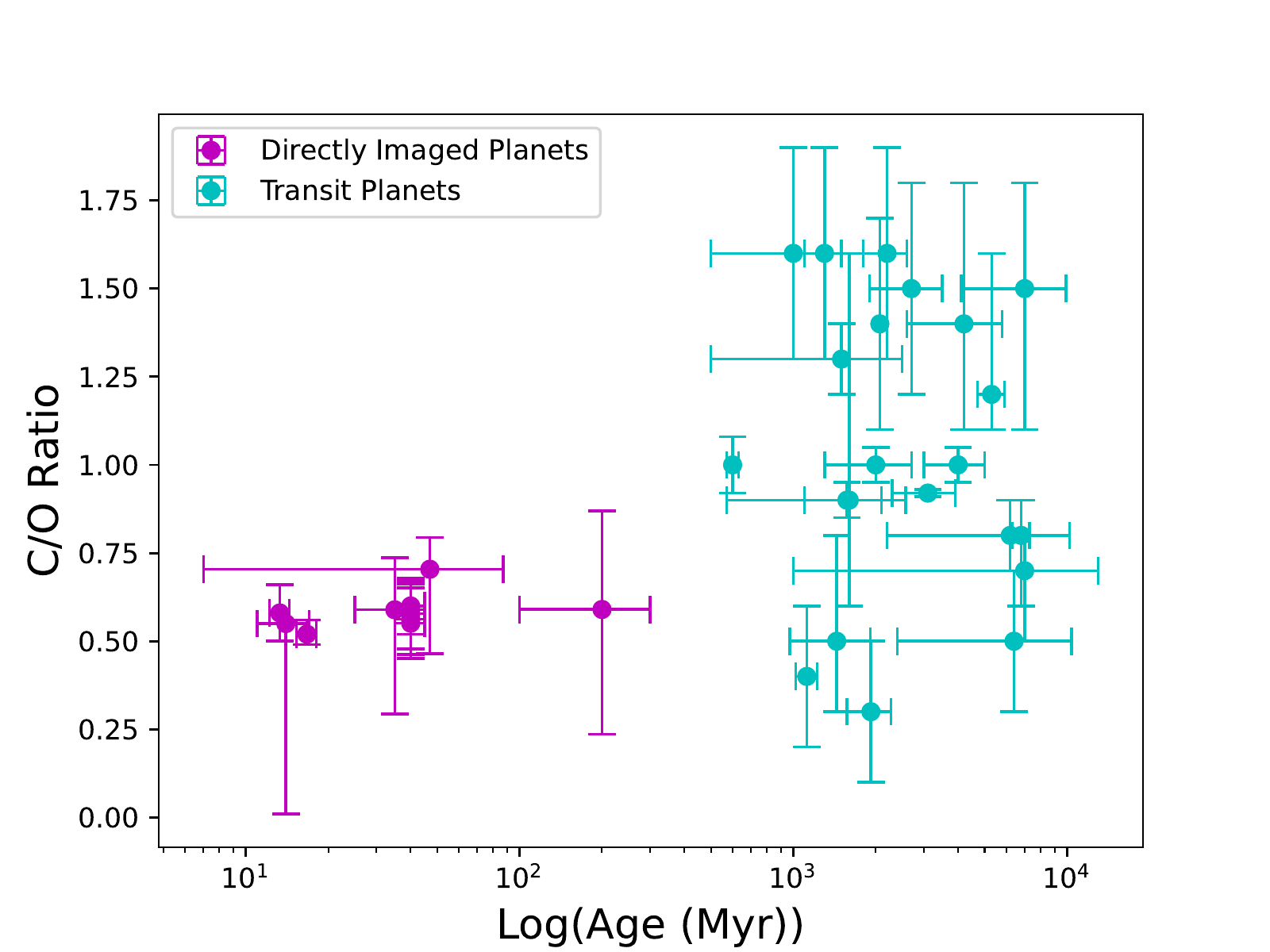}
\caption{Population of 25 retrieval exoplanets in blue and 9 directly imaged planets in magenta. Here are their C/O ratios plotted against the system age in Myr. This illustrates the differences in the two populations, with the directly imaged planets being younger because they are more luminous and more easily detectable with direct imaging. Even with this distinction, the correlation most likely is not with age, but with either companion mass or with semimajor axis.}
\label{fig:comp_age_c/o}
\end{figure*}

Table \ref{tab:megatable} shows all compiled values of C/O ratio for both transiting and directly imaged planets, including our new addition of HD 284149 AB b.  Other system parameters are included as well.  We explored C/O versus each of those parameters, noting whether the measurements are with transit or direct spectroscopy.  The results of this analysis are shown in Figures \ref{fig:comp_mass_c/o}--\ref{fig:comp_age_c/o}.  We see a trend when plotting companion mass against C/O ratio, shown in Figure \ref{fig:comp_mass_c/o}. All companions that are above 3--5\,$M_{\mathrm{Jup}}$ have C/O ratios around 0.7$\pm$0.2, and companions with masses less than 3--5\,$M_{\mathrm{Jup}}$ exhibit a wider spread, with C/O ratios of up to 1.6. 

\begin{figure*}
\centering
  \includegraphics[width=\textwidth]{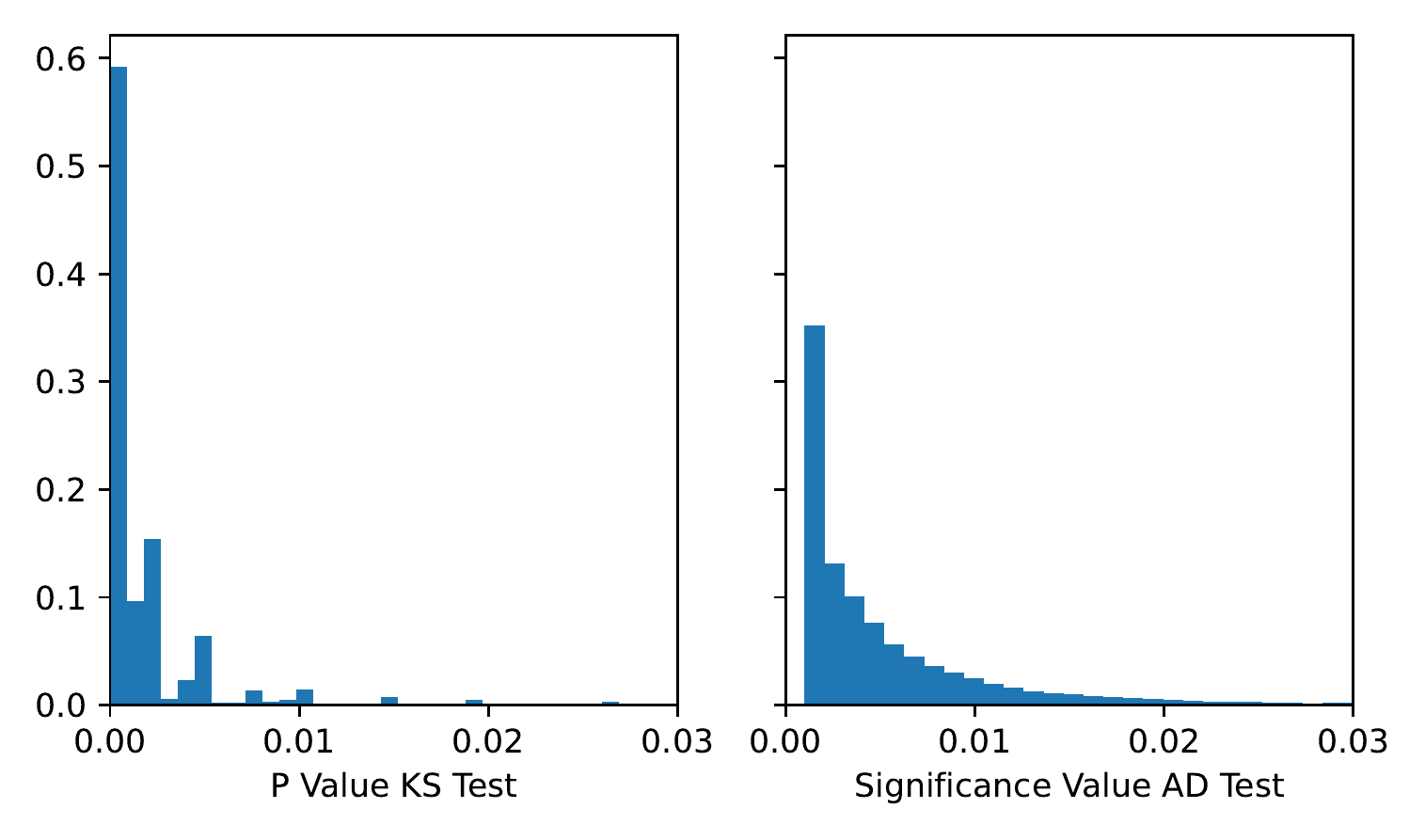}
\caption{A histogram of p values from the KS Test and significance values from the AD Test. The results seem to show that both p values and significance values are less than 1\% indicating that the two populations were distinct and not from the same underlying population.}
\label{fig:ks_test_ad_test}
\end{figure*}

To verify this distinction in C/O ratio around 3--5 M$_{\mathrm{Jup}}$, we performed a few statistical tests. We conducted a Kolmogorov–Smirnov Test (KS Test), and an  Anderson-Darling Test (AD Test) because they are well-established tests of similarity to determine if two distributions come from the same parent distribution. We first randomly sampled 100,000 values from each parameter (C/O ratio and companion mass) using a Gaussian distribution centered on the central value with a width defined by the uncertainty. For non-symmetric error bars we used a triangle function, centered on the central value with two widths of upper and lower error ranges. We chose 4 M$_{\mathrm{Jup}}$ as the distinguishing mass, the central value between 3 and 5\,$M_{\mathrm{Jup}}$. We then split the sample into two components, those with mass below 4\,$M_{\mathrm{Jup}}$ and the other equal to or greater 4\,$M_{\mathrm{Jup}}$. The two subsamples, with randomly sampled C/O ratios and corresponding randomly sampled masses were then put through the KS Test using the scipy.stats.ks$\_$2samp Python package and the AD Test using the scipy.stats.anderson$\_$ksamp Python package. This generated 100,000 p values and significance values from each test. We show the resulting distributions in Figure \ref{fig:ks_test_ad_test}. The results show that both p values and significance values are less than 1\%, indicating that the two populations were distinct and statistically unlikely drawn from the same underlying population. We performed this same test for a cut off at 3\,$M_{\mathrm{Jup}}$ and 5\,$M_{\mathrm{Jup}}$, with results revealing that the p values and significance values are less than 1\%. We also performed this test for just the transiting planets with a 4 M$_{\mathrm{Jup}}$ cut off, and found that the peak p values and significance values were less than 1\%.

The values derived for C/O could be subject to additional systematics that are unaccounted for in the error bars provided in the literature.  In particular, \citet{changeat2022} show that different C/O ratios are derived in retrievals depending on which datasets are included in the fits.  In order to try to account for these systematics, we performed the statistical similarity tests four more times, inflating the transit/eclipse C/O ratio errors by 25\%, 50\%, 75\%, and 100\% respectively.  We chose to inflate the transiting planet error bars but not the directly imaged planets because the former have much smaller uncertainties than the directly imaged planets, and are likely driving the visual distinction between the two populations of planets. For these four tests with inflated error bars, we found the same result with p values and significance values that were less than 1\%, indicating the populations are statistically distinct. 

The transit/eclipse and directly imaged planets overlap in mass, but not in separation, due to detection sensitivities, making mass a more interesting parameter to explore.  However, we do note that Figure \ref{fig:comp_sep_c/o} shows a visual difference in C/O ratio with separation, highlighting that all the directly imaged planet C/O ratios are very consistent with each other thus far.  Even those this visual difference may be driven entirely by detection biases, we also ran these two tests for C/O ratio versus separation (au) and obtained a similar result with both p values and significance values less than 1\%.  

\section{Discussion} \label{sec:discussion3}
\subsection{Implications for HD 284149 AB b} 
HD 284149 AB b is an interesting case for detailed atmospheric characterization because of its wide separation, youth, and brightness. Such objects that are easier to observe can provide insights into higher contrast systems for which it is difficult to obtain moderate resolution spectra. Our results of $T_\mathrm{eff} = 2502~K$, with a range of 2291--2624~K, $\log g=4.52$, with a range of 4.38--4.91, and [M/H] = 0.37, with a range of 0.10--0.55 agree with the results from \cite{bonavita2014,bonavita2017} that posit HD 284149 AB b is likely a brown dwarf mass object, in a young system.  We compared our results to \cite{baraffe2015} evolutionary models assuming an age of 25 Myr and found a corresponding mass range of 27--38\,$M_{\mathrm{Jup}}$, which is  consistent with 26 $\pm$ 3\,$M_{\mathrm{Jup}}$ from \cite{bonavita2017}.

Formation of objects in the same mass regime as gas giant planets and low mass brown dwarfs is of considerable interest because formation models have trouble producing these types of objects. Formation diagnostics such as C/O ratio can provide context as to where and potentially how these substellar objects have formed. HD 284149 AB b shares properties with imaged objects like AB Pic b \citep{chauvin2005} and ROXs 42Bb \citep{kraus2014}, which place HD 284149 AB b between the planetary mass object regime and the lowest mass brown dwarfs imaged to date. HD 284149 AB b represents a challenge in our understanding of formation of low-mass companions at extremely wide separations. The low mass ratio of this system might point towards a planet-like formation scenario such as core/pebble accretion, but its estimated mass is quite high, which could suggest a star-like rapid formation process such as gravitational instability. 

Measuring an approximately solar C/O ratio of HD 284149 AB b provides one possible diagnostic for understanding how this object formed. This result points to a very rapid formation process, potentially through either gravitational instability or common gravitational collapse similar to a binary star system.  However, this diagnostic involves a comparison to the host star in order to draw definitive conclusions about formation. We do not have a C/O ratio of the host star because the data available were low resolution only. The slightly elevated metallicity could shed some light on formation, but host star metallicity is still required. Until atmospheric characterization is conducted on HD 284149 A, we can only conclude that the evidence points to roughly similar values for the host star and the companion if the star has similar abundances to Solar.

HD 284149 AB b now represents a ninth directly imaged substellar companion, in addition to the four HR 8799 planets \citep{konopacky2013,barman2015,molliere2020,ruffio2021}, $\kappa$ And b \citep{wilcomb2020}, VHS 1256 b \citep{hoch2022}, HIP 65426 b \citep{petrus2021}, TYC 8998-760-1 b (YSES-1 b \citealt{zhang2021}), and AB Pic b \citep{palma-bifani2022} with an approximately solar C/O ratio.  The diagnostic did not produce the elevated C/O that would have been a strong argument in favor of core/pebble accretion. Although the scenario certainly cannot be ruled out given the uncertainties in the data and the range of possible C/O ratios predicted by core/pebble accretion models (e.g., \citealt{madhu2019}). 

\subsection{Implications for the C/O Ratio Trend}
All of the directly imaged planets with C/O ratios have estimated masses of $>$4 M$_{Jup}$. \cite{santos2017} postulated that there may be two distinct populations of exoplanets that split at 4 M$_{\mathrm{Jup}}$ when comparing stellar metallicity [Fe/H] to companion mass M$_{\mathrm{Jup}}$.  When we look at C/O ratio versus a range of other system parameters, the only one that showed a potential pattern or correlation and was not caused by detection bias, was companion mass. This could point towards two distinct populations of exoplanets based on companion mass, rather than detection method. Interestingly, the split is around 4 M$_{\mathrm{Jup}}$, the same as in \cite{santos2017}.  \cite{schlaufman2018} also shows that objects with masses $\le$4 M$_{\mathrm{Jup}}$ orbit metal-rich solar-type stars, a property that may indicate a core/pebble accretion formation pathway. The results of our KS test and AD test show that these populations are distinct populations and could possibly point to two different formation mechanisms.  We note that in order to confirm this trend, we must obtain more C/O ratios for lower mass directly imaged planets.  There are a few directly imaged planets with estimated masses below 4 M$_{Jup}$, such as 51 Eri b, \cite{macintosh2015}, but they have not yet had their C/O ratios measured.  Though we tried to obtain a spectrum 51 Eri b with OSIRIS, it remains beyond the reach of that instrument's capabilities.  Such measurements, which might be possible with the newer generation of instruments such as KPIC on Keck \cite{delorme2021}, may reveal whether the trend with mass holds. 

Additionally, there are some outliers in the samples probed in Figure \ref{fig:comp_mass_c/o}. In particular there is a large spread in C/O in the planets under 4 M$_{\mathrm{Jup}}$, all of which are computed with transit spectroscopy. This spread could have a physical meaning, or it could mean better constraints on C/O ratios are needed.  Indeed, the large spread in C/O ratio demonstrated in \cite{changeat2022} based on the dataset used for retrieval highlights the potential for systematic errors using currently available transit spectroscopy. \textit{JWST}, and future facilities such as \textit{ARIEL} and Extremely Large Telescopes (ELTs) will obtain tighter constraints on atmospheric species that trace the C/O value and could narrow this spread. 

\subsection{Implications for the C/O ratio as a Formation Tracer}
The C/O ratio as a formation tracer may require a deeper understanding of the evolution of the protoplanetary disk. There are other mechanisms that can impact the C/O ratio that should be studied before connecting composition to formation of transiting and directly imaged planets. Migration paired with core/pebble accretion has been postulated to explain the final location and atmospheric composition of planets, but including the chemical evolution of the protoplanetary disk could reveal that less migration is needed to explain the chemical properties of the atmospheres \citep{molliere2022}.  The population of Hot Jupiter's studied by \citet{changeat2022} almost certainly underwent significant migration, which could play an important role in the evolution of their C/O ratios. When considering pebble accretion, the drift, evaporation, and accretion of pebbles reproduces planetary C/O values, but it is uncertain whether it can reproduce the high atmospheric metallicities that have been seen in some directly imaged planets such as HR 8799 e \citep{molliere2020}. Therefore, it is important to have an understanding of the chemical composition and evolution of the protoplanetary disk to understand the composition of the disk gas and solids that ultimately build the planets we detect. Measuring the C/O ratio of planets is still important to highlight trends and dichotomies in the population of exoplanets, but tracing a formation pathway with these measurements require better constraints on elemental abundances and more detailed evolutionary models and/or observations of the disk composition \citep{vandermarel2021}. 

\section{Conclusions}\label{sec:conclusions}

The C/O ratio is thought to be a tracer of planet formation. In this work, we presented a statistical analysis of the published C/O ratios of directly imaged exoplanets and compared them with an existing characterized sample of transiting exoplanets.

Using moderate-resolution spectroscopy, we have greatly expanded our knowledge of directly imaged companions by measuring their atmospheric parameters and C/O ratios. Here, we found HD 284149 AB b to have T$_\mathrm{eff} = 2502~K$, with a range of 2291--2624~K, $\log g=4.52$, with a range of 4.38--4.91, and [M/H] = 0.37, with a range of 0.10--0.55 which is in agreement with the results from \cite{bonavita2014,bonavita2017}. We measure the C/O ratio to be 0.589$^{+0.148}_{-0.295}$. Without the ratio for the host star, we cannot conduct a complete formation tracer analysis, but an approximately solar C/O ratio could point towards a rapid formation process such as gravitational instability or common collapse. 

After obtaining the C/O ratio for HD 284149 AB b, we add this measurement to the growing list of measured C/O ratios for directly imaged companions and compare these ratios to C/O ratios measured by \cite{changeat2022} for a sample of 25 ``Hot Jupiters". We analyze C/O ratios versus various system parameters as shown in Figures \ref{fig:comp_mass_c/o}--\ref{fig:comp_age_c/o} and see a trend when looking at Companion Mass ($M_{\mathrm{Jup}}$) versus C/O ratio, where there are two distinct groups of companions separated at 4\,$M_{\mathrm{Jup}}$. We conduct a KS Test and AD Test to show that these groups of objects do not come from the same underlying population. This result could reveal that there are two distinct groups of companions that form through different methods.

Population studies of exoplanets and their formation pathways still remain difficult due to detection biases from each of the discovery methods. C/O ratio measurements for both transiting exoplanets and directly imaged exoplanets are also very difficult to measure using ground-based observatories. However, our imaging spectroscopy survey of exoplanetary atmospheres with Keck/OSIRIS has provided six out of the nine measured C/O ratios for directly imaged companions. To increase the number of C/O ratios measured, and improve the constraints on these measurements, space-based observatories such as \textit{JWST} will provide increased wavelength coverage and higher signal-to-noise data for both transiting and directly imaged companions. The work presented here will pave the way for future studies based on the next generation of space telescopes, high resolution spectrographs, and ELTs. 

\section{Acknowledgments}
The authors thank the anonymous referee for aiding in the publication of this work. The authors would like to thank Randy Campbell, Heather Hershley, and Tony Connors for their support in obtaining these observations. K.K.W.H., Q.M.K, and T.S.B acknowledge support by the National Aeronautics and Space Administration under Grants/Contracts/Agreements No.NNX17AB63G and  80NSSC21K0573 issued through the Astrophysics Division of the Science Mission Directorate.  T.S.B. acknowledges support by the National Science Foundation under Grant No. 1614492. Any opinions, findings, conclusions, and/or recommendations expressed in this paper are those of the author(s) and do not reflect the views of the National Aeronautics and Space Administration. The data presented herein were obtained at the W. M. Keck Observatory, which is operated as a scientific partnership among the California Institute of Technology, the University of California, and the National Aeronautics and Space Administration. The W. M. Keck Observatory was made possible by the financial support of the W. M. Keck Foundation. The authors wish to acknowledge the significant cultural role that the summit of Maunakea has always had within the indigenous Hawaiian community. The author(s) are extremely fortunate to conduct observations from this mountain. Portions of this work were conducted at the University of California, San Diego, which was built on the unceded territory of the Kumeyaay Nation, whose people continue to maintain their political sovereignty and cultural traditions as vital members of the San Diego community. 

\facilities{Keck/OSIRIS}

\software{\textit{emcee} \citep{foreman-mackey2013}, \textit{SMART} \citep{hsu2021_smart,hsu2021}, \texttt{orvara} \citep{2021AJ....162..186B}}

\bibliography{bibliography}
\end{document}